\documentclass[sigconf]{acmart}

\AtBeginDocument{%
 }

\setcopyright{none}

\acmJournal{JACM}
\acmVolume{37}
\acmNumber{4}
\acmArticle{111}
\acmMonth{8}

\copyrightyear{2025}
\acmYear{2025}
\setcopyright{acmlicensed}\acmConference[FAccT '25]{The 2025 ACM Conference on Fairness, Accountability, and Transparency}{June 23--26, 2025}{Athens, Greece}
\acmBooktitle{The 2025 ACM Conference on Fairness, Accountability, and Transparency (FAccT '25), June 23--26, 2025, Athens, Greece}
\acmDOI{10.1145/3715275.3732109}
\acmISBN{979-8-4007-1482-5/2025/06}

\author{Dan Bateyko}

\affiliation{%
 \institution{Cornell University}
 \city{Ithaca}
 \state{NY}
 \country{USA}
}
\email{drb348@cornell.edu}
\author{Karen Levy}

\affiliation{%
 \institution{Cornell University}
 \city{Ithaca}
 \state{NY}
 \country{USA}
}
\email{karen.levy@cornell.edu}

\begin{document}

\title{One Bad NOFO? AI Governance in Federal Grantmaking}

\begin{abstract}
 Much scholarship considers how U.S. federal agencies govern artificial intelligence (AI) through rulemaking and their own internal use policies. But agencies have an overlooked AI governance role: setting discretionary grant policy when directing billions of dollars in federal financial assistance. These dollars enable state and local entities to study, create, and use AI. This funding not only goes to dedicated AI programs, but also to grantees using AI in the course of meeting their routine grant objectives. As discretionary grantmakers, agencies guide and restrict what grant winners do—a hidden lever for AI governance. Agencies pull this lever by setting program objectives, judging criteria, and restrictions for AI use. Using a novel dataset of over 40,000 non-defense federal grant notices of funding opportunity (NOFOs) posted to Grants.gov between 2009 and 2024, we analyze how agencies regulate the use of AI by grantees. We select records mentioning AI and review their stated goals and requirements. We find agencies promoting AI in notice narratives, shaping adoption in ways other records of grant policy might fail to capture. Of the grant opportunities that mention AI, we find only a handful of AI-specific judging criteria or restrictions. This silence holds even when agencies fund AI uses in contexts affecting people’s rights and which, under an analogous federal procurement regime, would result in extra oversight. These findings recast grant notices as a site of AI policymaking---albeit one that is developing out of step with other regulatory efforts and incomplete in its consideration of transparency, accountability, and privacy protections. The paper concludes by drawing lessons from AI procurement scholarship, while identifying distinct challenges in grantmaking that invite further study.
\end{abstract}


 \keywords{AI governance, federal grants, grant policy}

\maketitle

\section{Introduction}

In 2023, the Department of Housing and Urban Development (HUD) granted funds to public housing officials to curb crime through purchases such as security cameras for their properties. After HUD disbursed the funds, reporting shed light on grant winners who had repurposed the cameras to surveil their own residents---for example, using facial recognition to track tenants and video records to identify lease violations, justifying evictions. But when Congress members demanded that the funding agency intervene, the agency shrugged: housing officials had broken no rules. The grant program used to purchase the cameras banned neither extramural use of purchases nor artificial intelligence (AI) tools such as the camera’s facial recognition features \cite{EyesPoorCameras2023a}. With the grant's spirit broken but its letter intact, who had the power to intervene?

In AI policy, how U.S. federal agencies govern grants has gone overlooked. And understandably so. With grants, the government’s role would seem simple: give away money and step back. To be sure, agencies hold grant winners accountable for proper spending, creating significant reporting requirements. But while agencies conduct site visits and review grant reports \cite{cfrrulemonitoring}, they tend to focus more on fraud prevention than deep performance review \cite{CurrentStateGrants2023}. Although agencies can recover funds if award winners veer from stated goals, they rarely exercise this authority \cite{pasachoffAgencyEnforcementSpendingb,pasachoffFederalGrantRules2020a}. This perception of light-touch oversight may help explain why scholars across law \cite{nagleReviewEssentialsGrant1994,pasachoffAgencyEnforcementSpendingb,priceGrants2019}, political science \cite{berryPresidentDistributionFederal2010} and economics \cite{azoulayScientificGrantFunding2020} note that their disciplines gloss over the mechanics of how agencies draw up grant policies. 

But grantmaking agencies have one key power: they can plan ahead. After facing backlash over the misuse of its crime grants, HUD banned automated surveillance in future grants, patching what its prior policy permitted \cite{EyesPoorCameras2023a}. Discretionary grants---where agencies have authority to select recipients---enable federal support for a broad range of activities, from basic research to capacity building, carried out by non-federal entities like states, local governments, nonprofits, and universities \cite{granteligibilitygrantsgov}. 

For these grants, agencies can conduct their own merit review of applicants \cite{CFRUniform}, typically through competitions where they set standards, convene judges, and crown winners \cite{keeganFederalGrantsinAidAdministration}. Agencies can make award winners follow certain terms, such as receiving technical help, submitting to monitoring, and passing key reviews \cite{CFRConditions}. These pre-award policies matter because agencies rarely can add new rules after funding begins \cite{yehFederalGovernmentsAuthority2017}. Through grant planning, agencies prioritize who and what gets funded, and put winners on notice to use their award appropriately---all ways in which federal agencies can influence how grantees use AI.

This paper asks how agencies shape the use of AI technologies by grantees through agency-drawn priorities and conditions in discretionary grants. To do so, we examine a publicly available yet often overlooked data source: Notices of Funding Opportunity (“NOFOs”). These notices are more than simple announcements. NOFOs describe the purpose of a grant, specifying eligible and ineligible activities, review criteria, reporting requirements, and funding restrictions \cite{CFRUniform}. This article creates a novel dataset of these notices, analyzing them at scale for the first time to ask two key questions:

\begin{enumerate}
 \item Which discretionary grants promote AI, and how?
 \item What award-specific review criteria and restrictions do agencies set to regulate grantees’ use of AI?
\end{enumerate}
	
These questions gain urgency as rising federal AI funding thrusts more responsibility onto government agencies. Both the Biden and Trump administrations have pushed to expand AI grantmaking. In 2023, President Biden directed several agencies to prioritize funding for AI initiatives \cite{houseExecutiveOrderSafe2023}. During his first term, President Trump called for increased AI research funding \cite{voughtFiscalYearFY2020} and AI education through scholarships and training programs \cite{ExecutiveOrderMaintaining}. President Trump has continued this trajectory in his second term, issuing an Executive Order directing the Department of Education to support K-12 AI education through targeted grants \cite{trump2025ai}. Even in its 2026 presidential budget proposal, which aims to reduce federal spending, the Trump administration committed to sustained federal AI funding \cite{whitehouse2025skinnybudget}. Congressional interest, too, has intensified. In 2021, the National Security Commission on Artificial Intelligence proposed doubling each year the non-defense AI research and development budget: from \$8 billion in 2024 to \$16 billion in 2025 and ultimately \$32 billion by 2026 \cite{FinalReport}. A Senate AI working group endorsed this \$32 billion target in its own roadmap \cite{bipartisansenateaiworkinggroupDrivingUSInnovation2024}. That road map earmarks specific agencies, but AI-targeted funds might soon spread across the broad landscape of grantmaking agencies, who use grants to fulfill critical public missions---from healthcare to national security to diplomacy. And that discretionary spending is significant; in 2024, government-wide obligations totaled over \$239.5 billion \cite{ProjectGrantsFY2024}.
 
A wave of grant applicants proposing to use AI to meet social challenges might test agency capacity and expertise. Agencies can have limited visibility into grantee activities. When grantees successfully use AI to meet grant goals, agencies may be unable to identify and share their strategies with others. Similarly, agencies may fail to catch early warning signs of civil rights or public safety risks. The HUD example underscores how grant-funded AI tools---facial recognition, in HUD’s case---can be repurposed in ways policymakers never envisioned, echoing broader concerns raised about AI transparency and oversight. In these ways, unplanned AI uses can jeopardize basic agency responsibilities for grant programs.

This article proceeds as follows. We start by examining how Congress and the executive branch shape program goals in discretionary grants. Within this broader structure of grant governance, we find few policies related to AI, motivating us to examine individual funding programs. To understand how agencies establish policy at the program level, we analyze NOFOs. We identify and assess how agencies plan for AI use through their notices, gathering 40,514 NOFOs published from 2009 to 2024 on Grants.gov. We search these NOFOs for AI keywords sourced from an authoritative federal definition of AI and manually review those matches. We classify grant awards as either directly funding AI, such as those for research, development, and promotion of AI, or indirectly funding AI, such as those encouraging grantees to use AI to achieve program goals. We draw attention to funding in particular high-impact contexts identified in the White House Office of Management and Budget's AI policy \cite{youngAdvancingGovernanceInnovation2024}. Across government, agencies seek out AI when making grants in potentially risky contexts, including in law enforcement, education, and healthcare. In identifying AI-related grants, we reason that NOFOs are a useful window not only into policy, but also spending: we uncover grants discussing AI that spending records—often lacking in clarity and completeness—fail to capture. Based on these findings, we argue that the grantmaking process merits closer scrutiny to improve transparency and mitigate AI risk scenarios. We conclude with lessons drawn from the related literature on public AI procurement governance, identifying opportunities for further study. 

We make the following contributions:

\begin{enumerate}
\item We conduct the first review of AI policy in discretionary grants, identifying that grantmakers in our study largely do not use their authority to set conditions on grantees' AI use. Indeed, only nine grant programs from our dataset include AI-specific review criteria or conditions.
\item We find that agencies promote AI beyond official merit criteria, in ways not captured in aggregated reporting. This finding indicates agencies are putting more money on the table to encourage AI use than previous measures would suggest. We identify 407 grant awards that fund AI and encourage grantees to use AI to meet grant goals.
\item We introduce a new approach for identifying grants potentially funding AI activities, uncovering examples and promising leads that other methods of auditing grant funding would miss. 
\end{enumerate}

\section{Sources of Grant Planning Policy}
In this section, we explain key laws and policies that regulate agency discretion in programmatic grant planning. Eloise Pasachoff provides an exemplary overview of grant conditions, which we review here \cite{pasachoffExecutiveBranchControl2022}. Federal grant policy spans multiple tiers: national laws, agency-wide guidance, and individual grant policies. According to Pasachoff, policies can include administrative conditions, such as mandatory disclosures, and programmatic conditions, such as those that specify the populations a grant must serve \cite{pasachoffAgencyEnforcementSpendingb}. We focus on programmatic conditions that affect grant outcomes. We find scant federal law or agency-wide guidance on AI oversight with respect to grantmaking, prompting our focus on individual grant policies described in NOFOs. 

Agency grantmaking power starts with explicit authorization from Congress. Typically, the statute that first establishes an agency refers to any grantmaking powers and determines the degree of agency discretion \cite{brownFederalAgencysAuthority1988}. At least 38 federal entities make discretionary grants \cite{ResourcesGrantseekers2023}. To guide grantmaking, the Federal Grant and Cooperative Agreement Act (FGCAA) defines what a grant is and, together with several other statutes, authorizes the Office of Management and Budget (OMB) to issue grant policies \cite{FederalGrantCooperative1978}. OMB sets the most prominent rules that affect grants in its Uniform Guidance \cite{UniformAdministrativeRequirements2013}. 

OMB’s Uniform Guidance governs grant planning in a section on “pre-federal award requirements.” There, OMB requires agencies to identify and communicate to recipients any relevant laws or requirements \cite{CFR200300,CFR200211}. Such federal mandates include federal “cross-substantive” or “cross-cutting” laws that apply across agencies, such as Executive Orders and federal statutes. These laws and policies act directly on agencies and recipients. For example, an Executive Order could prioritize certain goods or services for funding \cite{EnsuringFutureMade2021} or a law might ban certain purchases by grant recipients \cite{CFR200216}. Agency grant policies, established in agency-wide Terms and Conditions as well as tailored grant-to-grant, allow for similar control over grantee actions \cite{yehFederalGovernmentsAuthority2017}. Each agency reports any identified federal requirements for a grant program when listing the opportunity on the government clearinghouse, the System for Award Management (SAM) \cite{CFR200203}. These top-level policies offer the first clue as to how agencies manage AI in grants, which we turn to next.

\subsection{Conditions in National and Agency-Wide Policies}

This section reviews national and agency‑wide policies under the Biden Administration and, briefly, the early actions of the second Trump Administration that could influence programmatic grant conditions for AI.
 
Under the Biden administration, government initiatives for “trustworthy” or “responsible” AI framed the need for a “government-wide approach” to innovation and oversight \cite{houseExecutiveOrderSafe2023}. These initiatives included substantial changes in procurement contracting policy and attempts to modernize federal agencies through the use of AI \cite{youngM2410MEMORANDUMHEADS}. Despite this expansive interest, grantmaking received relatively little formal attention.
This forbearance may have been deliberate. In 2023, President Biden signed the Executive Order on Safe, Secure, and Trustworthy Artificial Intelligence (EO 14110), which established government-wide standards for AI safety \cite{houseExecutiveOrderSafe2023}. To advance AI governance consistent with EO 14110 and other relevant laws, OMB issued M-24-10, a memo to implement the order. That memo expressly exempted agencies from applying otherwise mandatory risk management practices to grants \cite{youngM2410MEMORANDUMHEADS}. Advocacy organizations, during a notice-and-comment period, urged that OMB apply the same rules to grantmaking, suggesting that the agency deliberately declined to do so \cite{danbateykoCDTCommentsOMB2023,ACLUEncouragesOMBa}. In contrast, the administration's parallel policy on biosecurity---another pillar of EO 14110---did apply risk controls to grant funding \cite{FRAMEWORKNUCLEICACID2024}. Other Biden-era OMB memoranda on AI, M-24-18 and M-21-06, similarly did not regulate agency grantmaking \cite{youngAdvancingResponsibleAcquisition2024,voughtM2106GuidanceRegulation2020}. Though OMB suggested grantmakers could voluntarily adopt AI risk practices consistent with policy \cite{youngM2410MEMORANDUMHEADS}, there is little evidence agencies did so.\footnote{In reviewing 22 agency plans to comply with the OMB AI guidance, we find only two mentions of grants or financial assistance \cite{USAgenciesPublish}. Both were cursory: HUD committed to engaging with “grantees” among other stakeholders when developing its AI approach, and NASA committed to building AI talent through university grants. It is possible that sub-agencies have begun setting grant guidelines following OMB’s AI guidance, but if broader policy existed, we would expect to see it cited in grant NOFOs. That said, at least one grant we identified outside our study's collection period showed voluntary compliance with OMB's policy: a January 2025 National Institute of Justice opportunity which funds researchers evaluating AI in the criminal legal system. In that grant, the agency requires applicants to identify potential safety and rights risks associated with their study's use of AI technologies, phraseology that draws on language from M-24-10 memo \cite{NIJFY25Research2025}.} 

Though still early in President Trump’s second term, recent actions show executive interest in controlling federal grantmaking. In early 2025, President Trump rescinded EO 14110 \cite{trump2025rescissions} and replaced it with EO 14,179, focused on reducing regulatory barriers to AI innovation \cite{trump2025barriersai}. Consistent with this shift, OMB withdrew its prior guidance and issued Memorandum M-25-21. Like the Biden-era order it replaced, the M-25-21 does not apply to grant programs \cite{vought2025}. As a result, federal grantmakers continue to operate without government-wide obligations for managing AI-related risks in grants.

That national and agency-wide policy are not active sites for AI governance does not preclude the possibility that agencies might be regulating AI uses on a program-by-program basis. To assess that possibility, we turn to grant-specific conditions found in NOFOs.

\subsection{NOFO Conditions}
All agencies must provide the same types of information in a NOFO, and they generally follow the same OMB template to do so. NOFOs include four required elements: the program description, the review criteria, the restrictions, and the eligibility requirements \cite{CFR200204}.

NOFOs begin with program descriptions. Beyond a basic outline, program descriptions can contain extensive context about the program’s creation, suggested ideas, and examples of desired solutions, including lists of successfully funded projects. Under review criteria, agencies disclose how they will judge applicants. These criteria can take the form of a point-based rubric grading applicants' alignment with program goals and policy preferences. Agencies will also describe their peer review or selection process \cite{AppendixPart200, mcgarityPeerReviewAwarding}. Under eligibility, agencies list types of eligible candidates as well as anything that may make a recipient or a project ineligible, including agency-set standards. Lastly, agencies can set special conditions for high-risk recipients, determined by a review of their history with compliance and potential to meet performance goals \cite{CFRConditions}.

To understand how these notices play out in practice, let us consider lasers. In “Point Cloud City,” a National Institute of Standards and Technology (NIST) grant, the agency aimed to fund a dataset of indoor maps created using a laser-based mapping technology \cite{PSIAPPointCloud2018}. NIST limited funding to state and local governments and their public safety organizations. In describing the opportunity, the agency set expectations that applicants explain how they plan to safely use lasers and notify building occupants when creating their maps. Later, in a section detailing how it will evaluate applicants, the agency stated it awards points for technical merit, defined, in part, as “the extent to which the proposal demonstrates a clear understanding of the program goals...and likely to achieve the NOFO’s stated objectives” \cite{PSIAPPointCloud2018}. The goals and standards in the program’s description are not referenced again in the review criteria, leaving applicants to triangulate them. As we turn to selecting a dataset to understand grant programmatic policy, we argue that no aggregating data source gathers these expectations, motivating us to review NOFOs.

\section{Methods}
Having described the layout of a NOFO, we now turn to a dataset of these documents to see if and how agencies specify AI priorities and conditions. This kind of thematic analysis of grant notices has proved successful in contexts evaluating individual agencies for their evidence-based criteria \cite{lee-eastonUtilizationEvidencebasedIntervention2022a}, climate justice efforts \cite{arkhurstRealizingJustice40Addressing2024}, and rhetorical devices \cite{millarTrendsUsePromotional2022}. Such studies focus on either single agencies or rely on short summaries to select records for review. Our method expands on these previous efforts by first creating a database of NOFOs across agencies, which we then search programmatically, enabling a systematic review across the federal grantmaking landscape.

\subsection{Data Collection}
We gather a dataset of NOFOs from Grants.gov. Established in 2002 by a Presidential e-government initiative \cite{georgew.bushPresidentsManagementAgenda2002}, Grants.gov centralizes NOFO summaries for 38 grantmaking entities. These organizations include 26 agencies \cite{GrantMakingAgenciesGrantsgov} and 12 independent federal agencies, executive branch offices, and commissions \cite{OtherGrantMakingAgencies}. Grants.gov offers archives of these summaries and metadata \cite{XMLExtractGrantsgov} and hosts web pages for each NOFO record where agencies can attach the full text of the notice. When agencies provide the full text, they do so as zip file attachments to Grants.gov web pages. To gather these zip files, we conducted three crawls of Grants.gov, completed in September 2023, April 2024, and January 2025. 

We take several filtering steps to narrow the dataset for our study. We filter our time period to 2009-2024 and remove from consideration any supporting documents, like appendices, FAQs, and third party reports contained in the attachments. Additionally, we filter out funding opportunities from the intelligence and defense agencies. We do so in part because these agencies have separate, existing guidelines, leading OMB to exempt them from AI procurement policy \cite{youngM2410MEMORANDUMHEADS}. We include cooperative agreements, a related funding instrument. While similar to grants, cooperative agreements allow for agencies to work with grant winners in developing solutions, potentially enabling support like technical assistance or oversight of recipients’ grant activities during the award period. At the NOFO stage, agencies can leave the decision between choosing a cooperative agreement and grants open, informing our decision to include them in our analysis. In total, we successfully match 40,514 complete NOFOs to the Grants.gov summary records.\footnote{The dataset is available upon request from the first author.}

\subsection{Search Criteria}
To identify relevant NOFOs, we search for AI-related terms drawn from definitions found in both M-25-21 and the rescinded M-24-10. While neither memo applies directly to federal financial assistance, as the most analogous guidance available, agencies would likely reference the definitions and obligations in M-25-21 when developing their AI grant programs. As such, we use their AI-related terms to construct an inventory of AI-related grant opportunities. The specific keywords we identify and use are listed in Table~\ref{tab:ai_keywords}.

\begin{table}[h]
\centering
\begin{tabular}{p{.5\textwidth}}
\toprule
\textbf{AI Keywords} \\
\midrule
artificial intelligence, machine learning, deep learning, supervised learning, unsupervised learning, semi-supervised learning, reinforcement learning, transfer learning, generative AI, genAI, foundation model, AI model, model weight \\
\bottomrule
\end{tabular}
\caption{List of AI keywords found in OMB memos M-25-21 and M-24-10 and used in searching NOFO records.}
\label{tab:ai_keywords}
\end{table}

\begin{figure}[h]
\centering
\includegraphics[width=\linewidth]{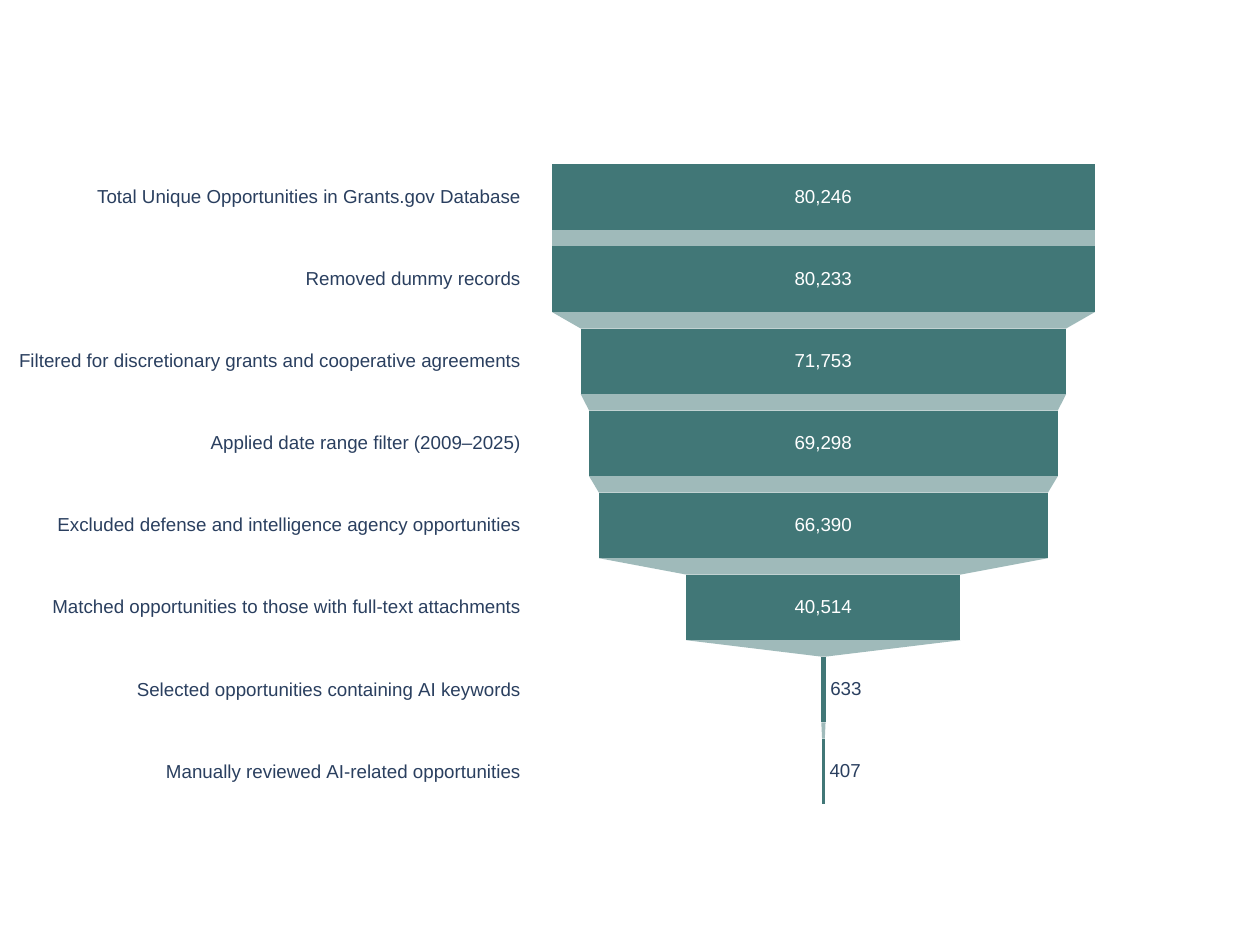}
\caption{A funnel chart showing the progressive filtering of notices in the Grants.gov database. The chart begins with 80,246 total opportunities and narrows down through stages such as removing dummy records, applying filters for discretionary grants and date ranges, excluding defense and intelligence agencies, and identifying AI-related opportunities through manual review, resulting in 407 AI-related notices.}
\Description{A funnel chart showing the progressive filtering of grant opportunities in the Grants.gov database. The chart begins with 80,246 total opportunities and narrows down through stages such as removing dummy records, applying filters for discretionary grants and date ranges, excluding defense and intelligence agencies, and identifying AI-related opportunities through manual review.}
\label{fig:funnel_chart}
\end{figure}

Our search yields 633 grant announcements with an AI keyword to review manually. During this process, we discard irrelevant matches such as those that do not contribute to the announcement (e.g. `machine learning' appearing in a boilerplate appendix; or a third-party’s AI use). After these steps, 407 opportunities remain to analyze qualitatively. We illustrate these filtering steps in Figure ~\ref{fig:funnel_chart}.

NOFOs in our dataset mention AI in ways that NOFO summaries and associated spending records do not capture. To confirm this finding, we examine two other sources: grant summaries from Grants.gov and spending records linked to a grant program’s assistance listings on USASpending.gov. We document this comparison in Appendix~\ref{sec:comparison}. Although mentions of AI in NOFOs do not necessarily materialize as AI development and use, this discrepancy suggests that the federally-funded AI footprint is underreported. 

\subsection{Qualitative Thematic Analysis}
We review the resulting set of 407 opportunities to: (1) classify how programs relate to AI, (2) identify programs that operate in contexts with potential risks to people's rights, and (3) determine whether agencies set AI-specific policies in their review criteria or restrictions. For each matching NOFO, we analyze the paragraph(s) in which an AI-related keyword appears. Using an inductive coding approach, we categorize the context in which the funding opportunity discusses AI, arriving at several recurring categories.

To identify potential risks to civil rights, liberties, and public health, we draw on a list of such use-cases from OMB memos. Both the current M-25-21 and rescinded M-24-10 memo outline purposes where, if an AI use will significantly influence the outcome of an agency decision, the use presumptively merits further oversight.\footnote{While the current guidance, M‑25‑21, retains most of the same enumerated use cases from the rescinded memo \cite{andrews2025omb}, it consolidates the previous memo's “safety-impacting” and “rights-impacting” uses under a single “high-impact AI” designation. Our analysis focuses exclusively on the rights dimension of that category, though we use “high-impact” and “rights-impacting” interchangeably for convenience.} Those purposes span a wide range of government activity, including government benefits, insurance, medicine, housing, immigration, law enforcement, and social media monitoring \cite{youngM2410MEMORANDUMHEADS}. Under current policy, procured AI uses that are designated as ``high-impact'' must undergo additional scrutiny, including an AI impact assessment of its cost and potential civil rights effects; pre-deployment testing; ongoing agency monitoring; and public consultation with affected members of the public \cite{vought2025}. Here, we take great caution. An announcement provides insufficient information to say with certainty where a proposed AI use is high-impact. For the purposes of our study, we simply document when the use case is in a context similar to those enumerated by OMB, indicating the need for further investigation. 

Lastly, we review each announcement to determine if the agency establishes specific AI-related policies. We do not record conditions which appear in our dataset and could affect AI use unless the agency indicates their applicability to AI (i.e. general data management policies). Finding few AI-specific restrictions that apply to grantees, we describe other ways agencies use NOFOs to promote the responsible use of AI. 

\subsection{Limitations}
Our dataset has variable coverage of NOFO documents within and across agencies. Though OMB requires agencies to document summary information about NOFOs on Grants.gov, uploading the full NOFO text is generally optional. Many agencies instead host full notices on their own platforms, providing only a link on Grants.gov. As a result, our findings understate the total volume of AI-related federal grant activity and may miss certain governance approaches used by agencies absent from our dataset. Figure~\ref{fig:agency_attachments} shows the variation in NOFO attachment availability across agencies and years. Further, our keywords do not capture the full range of AI-related terminology. Indeed, other federal policies, such as guidance from the Federal Chief Information Officers Council (CIO Council), have different AI definitions \cite{EO13960Artificial2023}. AI terminology has also changed over time; our search terms do not include such phrases as ``big data'' which may have served as proxies for similar emerging technologies over the years. That said, these keywords provide a starting point aligned with federal guidance, offering an example of how agencies might begin their own reviews of AI use cases. More specific terms absent from our keywords (e.g. ``large language model'') are frequently co-located with more general terms (e.g. ``artificial intelligence'') in our search. Additionally, the reliability of keyword searches depends heavily on the underlying document quality. Both the volume and clarity of NOFOs evolve over time as agencies change their transparency standards and plain language practices.

\section{Results}
\begin{figure}[htbp]
 \centering
 \includegraphics[width=1\linewidth]{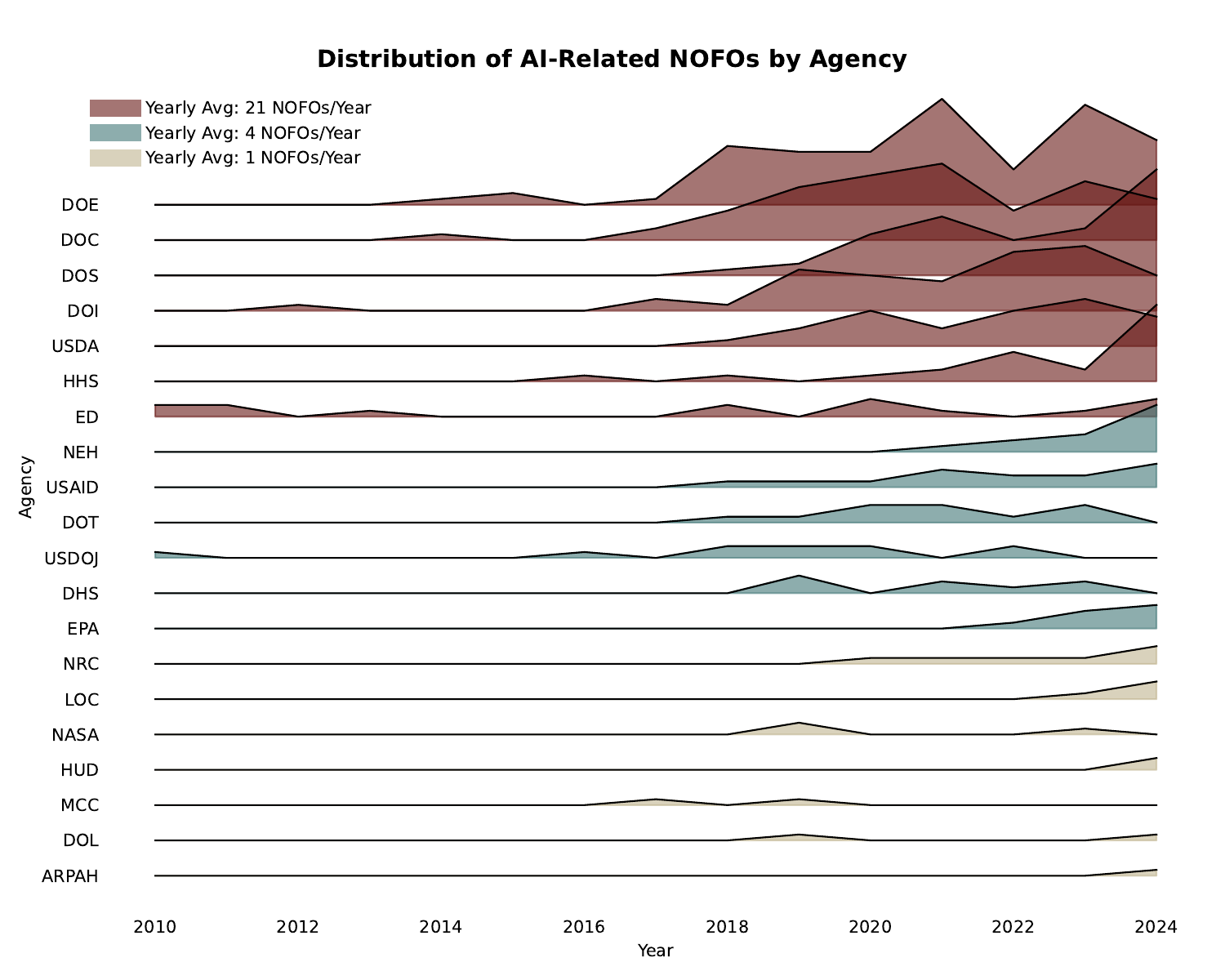}
 \caption{AI-related notice of funding opportunity trends by agency over time. We see a growing number of agencies that previously had not shown interest in AI prioritizing its research and use in their grant announcements starting from 2018. The full agency names matching these abbreviations are available in Appendix~\ref{fig:agencynames}.}
 \Description{A PDF figure showing the distribution of AI resources or technologies.}
 \label{fig:ai_distribution}
\end{figure}

Below, we share the results of our qualitative analysis. Through an iterative review, we classify the opportunities in our dataset into two broad categories: direct funding and indirect funding. Direct funding includes opportunities explicitly funding AI activities, such as those to develop an AI tool; conduct basic and applied research; or train people in working with AI. Indirect funding refers to opportunities that encourage grantees to use AI to achieve program goals without explicitly mandating it. We make this distinction on the premise that when agencies directly fund AI efforts, they are more likely to anticipate and influence AI uses. Under each broad category, we provide a description of subcategories and representative cases. Figure~\ref{fig:ai_distribution} illustrates the growth in AI-related NOFOs over time and across agencies, showing how interest has expanded to a wider range of departments since 2018.

AI keywords can appear throughout the NOFO, and AI-related policies can vary from firm conditions to high-level guidance. In previous years, OMB allowed agencies to scatter funding conditions throughout their announcements; in recent years, OMB has proposed reforms to organize conditions more systematically \cite{UniformGrantsGuidance2024}. As a result, many NOFOs lack a clear bright line for what ``counts'' as a requirement. In the statutory context, others have suggested that considerations---like that applicants ``take into account'' factors---can be read as rules \cite{hamburgerPurchasingSubmissionConditions2021}. The same applies here. Outside of enumerated conditions, we argue that \textit{de facto} conditions arise when program descriptions employ mandatory language (`must' or `shall') and merit review scores award points to applicants meeting program objectives. That agencies create informal conditions in notices poses challenges for applicants, agencies, and researchers. For applicants, the distinction between suggestions and requirements can blur, confusing their compliance expectations. Agencies, meanwhile, may lose track of the rules they impose, resulting in inconsistent enforcement. For researchers, these considerations---largely unexamined in the existing literature---make it difficult to identify all requirements at a glance, complicating the study of grantmaking.

The post-award funding amount of these awards can be challenging to calculate, for reasons discussed in Appendix~\ref{sec:comparison}. Agencies establish award ceilings and floors for grant programs, outlining broad ranges of the funding scope. Agencies in our dataset have average award ceilings ranging from \$250,000 to \$8 million for AI-related grant programs, with two notable exceptions: the Department of Commerce (\$60.3 million) and USAID (\$42.1 million). These outliers reflect the Department of Commerce's substantial investments in Regional Technology and Innovation hubs through the CHIPS and Science Act, and USAID's large-scale global health initiatives, some of which recommend AI technologies. 

\subsection{Direct Funding}
We examine direct funding for AI projects and classify opportunities into four categories.

In the first category of \textbf{AI as the Main Project Outcome}, grantmakers fund the creation or use of AI tools for immediate public needs. Grant announcements in this category typically describe the development and/or use of an AI tool as one of the grant opportunity’s expected outcomes or main priorities. These opportunities may include R\&D grants for which the list of deliverables includes developing and/or disseminating tangible AI-related products like training datasets or software, as well as ``extension'' activities that bring decision-support tools and findings directly to relevant stakeholders. AI can be the main focus of these awards or one of many program outcomes. Examples of AI tools in this category vary widely, from algorithms to locate protected species in the Gulf of Mexico to software that finds faults in residential heaters. We find 60 announcements in this category. 

To sponsor \textbf{AI as a Research Area}, agencies devise research proposals on problems specific to AI or its applications.\footnote{Notably, major researcher funders, such as the National Science Foundation and the National Institutes of Health, are absent from our dataset, likely because they host their own clearinghouses. As such, our sample  underrepresents basic research announcements.} Announcements in this category will describe the goal as research activities; this may include R\&D activities where the agency does not ask for tangible AI tools as deliverables. We find 124 announcements in this category. Examples in this category include a National Institute of Standards and Technology grant for public-safety first responders to utilize AI to identify common objects at an emergency site, and a 2020 Department of Transportation grant supporting research into AI for traffic management. Small business grants, like those from the Department of Education to stimulate R\&D from firms, can also fall under this category.

When \textbf{Promoting Educational \& Professional Development in AI}, grantmakers fund programs that teach AI concepts and build career pathways. We find 108 announcements in this category. An example grant opportunity supporting career development comes from the Department of Labor, which in 2019 sought to expand its apprenticeship programs to AI professions. In his 2019 Executive Order on Maintaining American Leadership in Artificial Intelligence, President Trump directed agencies to prioritize AI educational funding \cite{ExecutiveOrderMaintaining}; in several announcements in this category, agencies reference the Executive Order as an impetus for funding. Other examples include the Department of State grants funding the training of foreign officials and hosting, through embassies, educational programs on AI in countries like Croatia, Japan, and Macedonia.

To \textbf{Promote AI-Related Business Opportunities}, agencies recommend grantees collaborate with industry to execute their grant-funded missions. We find six grant announcements in this category. An example is a NASA and USAID-joint funded program to empower decision-makers with tools to act during natural disasters; in this grant, the funders note that private sector engagement is among the grant’s priorities, holding up past success among grantees who partnered with Google to use the company’s machine learning platforms for climate monitoring. The grant encourages prospective grantees to explore similar private-sector partnerships.


\subsection{Indirect Funding}
Grant announcements often \textbf{Suggest AI as a Tool for Project Objectives}, nudging recipients towards AI as `areas of opportunity,' `technologies of interest' or `attractive alternatives,' that can `unlock potential benefits.' In announcements, agencies take at least two different approaches: recommending AI tools because of previous success in the same field and recommending the tools because of successful uses in other contexts. These special encouragements appear when describing grant priorities, such as cleaning a dataset, predicting an outcome, or conducting basic research. Agencies make such suggestions throughout the announcement, though we find them most often in the program description. We find 109 announcements matching this category. Examples of these nudges include a Centers for Disease Control opportunity that suggests grantees use machine learning to identify spina bifida; a HUD opportunity that proposes machine learning to identify tenants at risk of eviction; and a National Park Service opportunity that strongly encourages applicants to create innovative machine learning tools for cultural preservation. 

\subsection{Grant Opportunities in Potential Rights-Impacting Contexts}
In our review of grant announcements, 25 are in contexts similar to those high-impact ones described in OMB’s M-25-21 memo. These contexts span law enforcement, education, and healthcare. Below we pull out illustrative examples.

In \textit{law enforcement contexts}, the National Institute of Justice (NIJ) and the Department of Homeland Security (DHS) make financial awards. In our review, we find several opportunities from these agencies soliciting the use of AI. NIJ, the research and development branch of the Department of Justice, provides funding in several contexts with inherent risks recognized by OMB policy, such as predicting recidivism and determinations related to parole. In a 2016 opportunity, NIJ aimed to improve the monitoring of people on probation or parole and under community supervision. The program envisioned funding recipients who would use AI and data analytics to predict re-offense risk and intervene. Another opportunity from NIJ funds research and interventions related to intimate partner violence while prioritizing machine-learning methods to investigate crimes. In these research programs, NIJ sets out that any datasets generated be shared with the agency. DHS, in its programs funded research for counterterrorism purposes, which may similarly fall under the presumptive high-risk category of identifying criminal suspects. 

In \textit{educational contexts}, the Department of Education offers funds to monitor student progress and outcomes, which may affect an individual's access to education. In one program, the Department’s Office of Special Education and Rehabilitative Services prioritizes applicants using AI to meet the needs of students with disabilities. Through the program, the agency seeks evidence-based approaches to identify learners’ needs and provide individualized instruction. 

In \textit{healthcare contexts}, several agencies provide funding to improve medical diagnoses through AI innovation. USAID offers funding soliciting the use of AI to ascertain patterns in social networks leading to identifying individuals at higher risk of contracting HIV---an example of flagging individuals for intervention that OMB might review as an AI use with a significant effect on human health and safety.

\subsection{Conditions and Considerations for AI Use}
We find nine opportunities that place explicit guardrails on the development or use of AI. The conditions impose two kinds of guardrails: to \textit{disclose the use of AI} and to \textit{discourage or prohibit AI use}.

In the case of disclosures, agencies will require applicants provide specific information to support their merit review. For example, we find one 2024 Department of Energy program requesting applicants to describe methods used and personnel responsible for carrying out AI-related activities. Such disclosures can include asking applicants to demonstrate awareness of AI risk policies: in 2024, the Department of Health and Human Services (HHS) issued an opportunity funding AI to improve healthcare data, asking applicants to show familiarity with safe and responsible AI principles as described by EO 14110 and HHS rules. Agencies can also require that AI prototypes be delivered to them at the end of the award for third-party auditing, as is the case in one 2018 NIJ opportunity to develop AI-based tools combating human trafficking. 

Concerns such as intellectual property rights, explainability, and capacity to evaluate proposals drive \textit{AI discouragements and prohibitions}. A Library of Congress program to help create educational materials to build student literacy restricts awardees from using generative AI to produce project materials, detailing examples of prohibited uses and listing example banned tools like ChatGPT, BARD, Dall-E, MidJourney, and Stable Diffusion. The grant also clarifies that AI used for routine tasks, like spellchecking or image editing, is permissible. Similarly, a grant program from the National Endowment for the Humanities discourages award winners from using generative AI to prepare grant-funded media projects. 

The need to explain algorithmic predictions also yields grant conditions: A 2024 Environmental Protection Agency opportunity to treat water pollution blocks from funding ``black box concepts'' like machine learning. Lastly, the State Department’s funding showcases an evolving policy on AI that has grown more prohibitive over time. In early announcements, the Bureau of Democracy, Human Rights, and Labor states AI-based projects are ``not typically considered competitive'' (i.e., unlikely to receive grant funding); in more recent policies, the Bureau strengthened its language to a restriction, stating programs ``must not'' focus on AI without a clear strategic application for human rights. In these announcements, the Bureau also requires applicants using AI to comply with federal principles on the use of AI. In another State Department announcement, the agency restricts funding for AI-based projects aimed at developing American expertise in Eastern Europe. As part of its reasoning, the Department states that the agency does not have an evaluation expert to assess AI proposals, pointing prospective applicants elsewhere.

Beyond restrictions, agencies signal the importance of responsible AI in other ways, including \textit{flagging AI-related equity concerns} and \textit{investing in measuring outcomes associated with AI}. The Department of Homeland Security and USAID highlight AI risks in program statements. USAID also includes links to resources on fairness and AI, recommending applicants consider those materials in order to submit a competitive proposal. Other agencies fund studies of AI’s ethical and safety-related issues. These include a 2024 State Department grant to mitigate AI-enabled chemical design risks; a 2023 National Endowment for the Humanities grant to study the ethical implications of AI; and a 2019 Department of Homeland Security grant asking applicants to propose strategies for their Civil Rights office to oversee machine learning tools. 

\section{Discussion}
Our central finding---that federal agencies rarely use grant conditions to govern AI---persists across administrations, regardless of shifting grant policy priorities. Whether the focus is on ideological bias, national security, economic competitiveness, or fairness and accountability, the core issue remains: when agencies forgo grant conditions, they forgo a key tool for advancing their priorities. Although a growing number of agencies promote AI in increasingly diverse ways, very few set AI-specific conditions at the pre-award stage. When they do, the conditions are broad disclosures, discouragements, or bans, rather than domain-specific policies tuned to the grant's purpose. That so few agencies articulate specific AI policies is not necessarily surprising: an AI tool's role may be too limited to justify extra oversight. Whether agencies should implement AI policies early in the funding lifecycle is a complex policy decision. To anticipate grant policy opportunities and challenges, we turn to an analogous context: government procurement policy.

\subsection{Lessons from AI Procurement Literature}
That most of the underlying issues outlined in the literature in AI procurement should also apply to grant is intuitive. Agency procurement offices, like grant offices, are inspired by private sector successes, and may sometimes overlook the importance of adapting these tools to their specific government contexts \cite{hendersonAdsSequentialDecisionMaking2022a,rajiFallacyAIFunctionality2022}. Vendor marketing exacerbates this problem when over-promising what an AI-driven solution can do, creating a challenging task for agencies and award winners funding and purchasing AI tools \cite{quay-delavalleeFederalGovernmentPower2024,narayananAISnakeOil2024a}. 

Another shared challenge for grantmakers and procurement officials alike is monitoring AI deployment. The transparency of AI spending has emerged as a key AI governance concern \cite{whittlestoneWhyHowGovernments2021} in part because officials cannot impose conditions without knowing first what `counts' as AI. Definitional debates can delay consensus in policymaking \cite{levyAlgorithmsDecisionMakingPublic2021a, krafftDefiningAIPolicy2020} and federal bureaucracies have struggled to track their use. Reviewing an early federal effort to inventory AI, Lawrence, Cui, and Ho \cite{lawrenceBureaucraticChallengeAI2023b} found that nearly half of agencies failed to include known machine-learning use cases. Their analysis suggests several obstacles, including that bureaucrats lack adequate training and face resource constraints. Notably, upskilling efforts for federal employees may not extend to many grantmakers; the 2022 AI Training Act, for example, does not explicitly cover grant specialists \cite{sen.petersAITrainingAct2022}. 

However, the flexibility in grants, like in procurement, provides room for tailored oversight of AI deployments. Coglianese argues that contracting is advantageous because it allows public officials to tailor restrictions to each deployed AI tool \cite{coglianeseProcurementArtificialIntelligence2023a}. Yet applying programmatic restrictions is a policy decision, and one not always acknowledged as such. The recognition that bureaucratic decisions embed political and ethical choices into public AI uses has been hard-earned. Mulligan and Bamberger \cite{mulliganProcurementPolicyAdministrative2019} propose that procurement officials make policy choices about the purchase and use of machine learning tools through procurement processes. In response, they recommend bringing administrative law values, like reasoned explanation, to the technical process of procurement. Much like in procurement, grant management goes beyond simple procedure. In fact, since the 1980s, policymakers have called for a more administrative-law-style grantmaking process in recognition of this role \cite{ResolvingDisputesFederal,PeerReviewAwarda}. 

\subsection{Grant-Specific Challenges}
AI procurement studies provide lessons for grant policymaking, but grant management policies and practices have their own idiosyncrasies which require further investigation. First, agency discretion in choosing contracts versus grants can enable evasion of oversight by misclassifying spending \cite{ScanwellChallengingPropriety}. Reasonable ambiguity exists in classifying funding: agencies have noted that R\&D does not fall easily into either category \cite{brownFederalAgencysAuthority1988}. Courts do not review misclassified grants unless a plaintiff alleges some legal injury, but in those cases, judicial scrutiny has corrected agency misclassifications \cite{360TrainingcomIncUnited,CMSContractMgmt}. 

Regulatory burden and the scope of administrative deference, a source of regular debate, is particularly contentious in grantmaking. Agency rules may add compliance requirements that grant recipients struggle to meet. For example, in a Congressional hearing on discretionary grants, stakeholders testified that certifying compliance can exclude and delay applicants from applying \cite{DepartmentTransportationDiscretionary2024}. Critics of agency policy-setting in grants might suggest agencies overstep by imposing non-statutory requirements, while proponents reason that agency conditions serve as legitimate tools stemming from the necessary delegation of authority to agencies \cite{pasachoffAgencyEnforcementSpendingb}. 

Guidance and capacity constraints can arise in grants due to the delegation of oversight without sufficient guidance for award winners, as is exemplified in law enforcement grants. In analyzing discretionary grants to law enforcement from the Department of Homeland Security and the Department of Justice, Crump \cite{crumpSurveillancePolicyMaking2016a} finds that although the government provided incentives for purchasing surveillance equipment, it did not offer guidance on appropriate uses. As a result, Crump suggests, officials were unclear of their oversight responsibilities and failed to draft comprehensive data management plans. This delegation of oversight becomes challenging as grantmakers may lack the capacity relative to procurement to oversee grant conditions. Friedman, Harmon and Heydari, when describing efforts to reform conditions in policing grants, point out that enforcers are stretched in overseeing numerous recipients \cite{barryfriedmanFederalGovernmentsRole}. 

\subsection{Future Work}
Our analysis suggests several topics for study and important areas beyond our immediate scope.

First, agencies face questions about how to incorporate AI expertise into their grantmaking processes. One approach to investigating this issue might involve conducting fact-finding through Requests for Information (RFIs) posted to Grants.gov. For example, in 2021, USAID solicited feedback on how gender bias in AI manifests in low and middle-income countries \cite{GenderInequityArtificial2021}. Crowdsourcing through grant portals may be a tailored way for agencies to gain domain knowledge from potential grant applicants. More study is also needed on how to effectively staff peer reviews to identify and judge AI applications. The extent to which agencies manage reviewer bias and conflicts of interest remains debated, reflecting broader questions of how to achieve objectivity in grant review processes \cite{mcgarityPeerReviewAwarding, priceGrants2019}. Recent controversies, like Congressional representatives' allegations that the National Institute of Standards and Technology’s AI Safety research grants ran afoul of sound science \cite{ScienceCommitteeLeaders2023}, exemplify these disputes. Future work could examine how agencies determine when to staff review panels with external experts and how they balance technical assessments with other review criteria.

Grants are one of many financial assistance tools---other tools, such as prize competitions (``challenges''), raise similar innovation and oversight questions. Crawford and Wulkan \cite{alicrawfordFederalPrizeCompetitions2021} suggest prize competitions might spur AI innovation, a recommendation OMB has adopted previously in guidance \cite{youngAdvancingResponsibleAcquisition2024}. These challenges may also raise concerns for privacy and equity, much like the grants in our dataset. Jegede et al. \\cite{jegedeChallengeAcceptedCritique2023} find ``serious ethical and methodological issues'' in the National Institute of Justice’s recidivism forecasting challenge attributed to the agency’s failure to adopt participatory AI design practices. Scholars might study creative pairings of funding instruments. For example, agencies that lack the capacity to host prize competitions can fund grantees to run them on the agency’s behalf \cite{SearchResultsDetail2019}. Some grants in our study mixed prizes with grants, allowing the agency to fund separately two applicants: one with the most accurate forecasting model, and one, in the agency’s judgment, with the best proposal to implement the model. 

Finally, although we look at the federal level, grant policymaking among states deserves attention as well. For example, in 2024, the Illinois General Assembly proposed a change to the state’s procurement code that would require AI disclosures in grants \cite{IllinoisProcurementCode2024}. 

\section{Conclusion}
When grants are thought of as donations, as they often are, questions about their oversight fade from view. But they are not donations. They are agreements made to deliver a public benefit. Through grants, agencies can pilot new approaches, gather evidence to inform future decisionmaking, and improve public services. However, how agencies incentivize innovation or anticipate missteps and misconduct when funding emerging technologies like AI has remained largely unexplored. For a variety of reasons, we argue that policy setting in federal discretionary grantmaking has been an unappreciated site of AI governance study, and we draw overdue attention to it in this paper. 

Across federal agencies in our study, we find discretionary grant notices recommend AI as a goal or tool, but rarely imposing conditions on its use. These promotions are a channel for influencing AI use and adoption. By turning attention to grants, we echo the call from Mulligan and Bamberger \cite{mulliganProcurementPolicyAdministrative2019} for a ``public sector culture of algorithmic responsibility'' in which bureaucrats move beyond a procedural mindset to promote transparency, fairness, and accountability in AI systems. Based on our findings, we emphasize the need for federal agencies to approach the setting of grant policy with an eye on AI’s challenges and opportunities. 

\begin{acks}
 We are indebted to Elizabeth Laird for prompting the initial inquiry into the governance of grantmaking, and to staff at the Center for Democracy \& Technology, including Hannah Quay-de la Valle, Ridhi Shetty, and Quinn Anex-Ries, for their thoughtful contributions during an early phase of this paper. We thank the participants of the 2024 Privacy Law Scholars Conference, including Hilary Robinson, Richmond Wong, Kendra Albert, Amina Abdu, Christopher Morten, and Lance Mabry, for their thoughtful feedback on an early draft. Tracy Hamler Carrick at Cornell University's John S. Knight Institute for Writing in the Disciplines provided valuable writing support. This paper benefited from conversations with Amanda Levendowski, Emma Lurie, Sara Stadulis, and members of the Cornell Artificial Intelligence, Policy, and Practice initiative. We are grateful to the John D. and Catherine T. MacArthur Foundation and the Canadian Institute for Advanced Research for support.
\end{acks}

\bibliographystyle{ACM-Reference-Format}
\bibliography{sample-base}


\begin{thebibliography}{93}


\ifx \showCODEN    \undefined \def \showCODEN     #1{\unskip}     \fi
\ifx \showDOI      \undefined \def \showDOI       #1{#1}\fi
\ifx \showISBNx    \undefined \def \showISBNx     #1{\unskip}     \fi
\ifx \showISBNxiii \undefined \def \showISBNxiii  #1{\unskip}     \fi
\ifx \showISSN     \undefined \def \showISSN      #1{\unskip}     \fi
\ifx \showLCCN     \undefined \def \showLCCN      #1{\unskip}     \fi
\ifx \shownote     \undefined \def \shownote      #1{#1}          \fi
\ifx \showarticletitle \undefined \def \showarticletitle #1{#1}   \fi
\ifx \showURL      \undefined \def \showURL       {\relax}        \fi
\providecommand\bibfield[2]{#2}
\providecommand\bibinfo[2]{#2}
\providecommand\natexlab[1]{#1}
\providecommand\showeprint[2][]{arXiv:#2}

\bibitem[Dim({[n.\,d.]})]%
        {DimensionsAIMost}
 \bibinfo{year}{[n.\,d.]}\natexlab{}.
\newblock \bibinfo{title}{Dimensions AI}.
\newblock
\newblock
\urldef\tempurl%
\url{https://www.dimensions.ai/}
\showURL{%
\tempurl}


\bibitem[Fed({[n.\,d.]})]%
        {FederalAwardsSearch}
 \bibinfo{year}{[n.\,d.]}\natexlab{}.
\newblock \bibinfo{title}{Federal Awards, Search for ``Facial Recognition'' {\textbar} ``Cameras'' from Department of Housing and Urban Development (HUD)}.
\newblock
\newblock
\urldef\tempurl%
\url{https://www.usaspending.gov/search/?hash=bf433b0e8636216d3c0cda3e0e38b04a}
\showURL{%
\tempurl}


\bibitem[Gov({[n.\,d.]})]%
        {GoviniArkPlatform}
 \bibinfo{year}{[n.\,d.]}\natexlab{}.
\newblock \bibinfo{title}{Govini Ark}.
\newblock
\newblock
\urldef\tempurl%
\url{https://www.govini.com/products/ark}
\showURL{%
\tempurl}


\bibitem[Fed(1978)]%
        {FederalGrantCooperative1978}
 \bibinfo{year}{1978}\natexlab{}.
\newblock \bibinfo{title}{Federal Grant and Cooperative Agreement Act of 1977}.
\newblock
\newblock
\urldef\tempurl%
\url{https://www.govinfo.gov/content/pkg/STATUTE-92/pdf/STATUTE-92-Pg3.pdf}
\showURL{%
\tempurl}


\bibitem[sen(2006)]%
        {sen.coburnText2590109th2006}
 \bibinfo{year}{2006}\natexlab{}.
\newblock \bibinfo{title}{Federal Funding Accountability and Transparency Act of 2006}.
\newblock
\newblock
\urldef\tempurl%
\url{https://www.congress.gov/bill/109th-congress/senate-bill/2590/text}
\showURL{%
\tempurl}


\bibitem[360(2012)]%
        {360TrainingcomIncUnited}
 \bibinfo{year}{2012}\natexlab{}.
\newblock \bibinfo{title}{360Training.Com Inc. v. United States}.
\newblock
\newblock


\bibitem[Uni(2013)]%
        {UniformAdministrativeRequirements2013}
 \bibinfo{year}{2013}\natexlab{}.
\newblock \bibinfo{title}{Uniform Administrative Requirements, Cost Principles, and Audit Requirements for Federal Awards}.
\newblock
\newblock
\urldef\tempurl%
\url{https://www.federalregister.gov/documents/2013/12/26/2013-30465/uniform-administrative-requirements-cost-principles-and-audit-requirements-for-federal-awards}
\showURL{%
\tempurl}


\bibitem[CMS(2014)]%
        {CMSContractMgmt}
 \bibinfo{year}{2014}\natexlab{}.
\newblock \bibinfo{title}{CMS Contract Mgmt. Servs. v. Mass. Hous. Fin. Agency}.
\newblock
\newblock


\bibitem[Exe(2019)]%
        {ExecutiveOrderMaintaining}
 \bibinfo{year}{2019}\natexlab{}.
\newblock \bibinfo{title}{Executive Order on Maintaining American Leadership in Artificial Intelligence -- The White House}.
\newblock
\newblock
\urldef\tempurl%
\url{https://trumpwhitehouse.archives.gov/presidential-actions/executive-order-maintaining-american-leadership-artificial-intelligence/}
\showURL{%
\tempurl}


\bibitem[Sea(2019)]%
        {SearchResultsDetail2019}
 \bibinfo{year}{2019}\natexlab{}.
\newblock \bibinfo{title}{NIST Public Safety Innovation Accelerator Program: First Responder 3D Indoor Tracking Prize}.
\newblock
\newblock
\urldef\tempurl%
\url{https://grants.gov/search-results-detail/321433}
\showURL{%
\tempurl}


\bibitem[Ens(2021)]%
        {EnsuringFutureMade2021}
 \bibinfo{year}{2021}\natexlab{}.
\newblock \bibinfo{title}{Ensuring the Future Is Made in All of America by All of America's Workers}.
\newblock
\newblock
\urldef\tempurl%
\url{https://www.federalregister.gov/documents/2021/01/28/2021-02038/ensuring-the-future-is-made-in-all-of-america-by-all-of-americas-workers}
\showURL{%
\tempurl}


\bibitem[sen(2022)]%
        {sen.petersAITrainingAct2022}
 \bibinfo{year}{2022}\natexlab{}.
\newblock \bibinfo{title}{AI Training Act}.
\newblock
\newblock
\urldef\tempurl%
\url{https://www.congress.gov/bill/117th-congress/senate-bill/2551/text}
\showURL{%
\tempurl}


\bibitem[EO1(2023)]%
        {EO13960Artificial2023}
 \bibinfo{year}{2023}\natexlab{}.
\newblock \bibinfo{booktitle}{\emph{EO 13960 Artificial Intelligence (AI) Use Case Inventories: Guidance for Creating Agency Inventories of AI Use Cases Per EO 13960}}.
\newblock \bibinfo{type}{{T}echnical {R}eport}. \bibinfo{institution}{Federal Chief Information Officers (CIO) Council}.
\newblock
\urldef\tempurl%
\url{https://www.cio.gov/assets/resources/2023-Guidance-for-AI-Use-Case-Inventories.pdf}
\showURL{%
\tempurl}


\bibitem[Res(2023)]%
        {ResourcesGrantseekers2023}
 \bibinfo{year}{2023}\natexlab{}.
\newblock \bibinfo{booktitle}{\emph{Resources for Grantseekers}}.
\newblock \bibinfo{type}{{T}echnical {R}eport}. \bibinfo{institution}{Congressional Research Service}.
\newblock
\urldef\tempurl%
\url{https://crsreports.congress.gov/product/pdf/RL/RL34012}
\showURL{%
\tempurl}


\bibitem[Sci(2023)]%
        {ScienceCommitteeLeaders2023}
 \bibinfo{year}{2023}\natexlab{}.
\newblock \bibinfo{title}{Science Committee Leaders Stress Importance of Diligence in NIST AI Safety Research Funding}.
\newblock
\newblock
\urldef\tempurl%
\url{https://science.house.gov/2023/12/science-committee-leaders-stress-importance-of-diligence-in-nist-ai-safety-research-funding}
\showURL{%
\tempurl}


\bibitem[CFR(2024a)]%
        {CFRUniform}
 \bibinfo{year}{2024}\natexlab{a}.
\newblock \bibinfo{title}{{2 CFR § 200}}.
\newblock \bibinfo{howpublished}{\url{https://www.ecfr.gov/current/title-2/part-200}}.
\newblock
\newblock
\shownote{Uniform Administrative Requirements, Cost Principles, and Audit Requirements for Federal Awards}.


\bibitem[CFR(2024b)]%
        {CFR200203}
 \bibinfo{year}{2024}\natexlab{b}.
\newblock \bibinfo{title}{{2~CFR~§~200.203}}.
\newblock \bibinfo{howpublished}{\url{https://www.ecfr.gov/current/title-2/part-200/section-200.203}}.
\newblock
\newblock
\shownote{Requirement to Provide Public Notice of Federal Financial Assistance Programs}.


\bibitem[CFR(2024b)]%
        {CFR200204}
 \bibinfo{year}{2024}\natexlab{b}.
\newblock \bibinfo{title}{2 CFR § 200.204}.
\newblock
\newblock
\urldef\tempurl%
\url{https://www.ecfr.gov/on/2024-10-01/title-2/part-200/section-200.204}
\showURL{%
\tempurl}
\newblock
\shownote{Notices of Funding Opportunities}.


\bibitem[CFR(2024a)]%
        {CFRConditions}
 \bibinfo{year}{2024}\natexlab{a}.
\newblock \bibinfo{title}{{2~CFR~§~200.208}}.
\newblock \bibinfo{howpublished}{\url{https://www.ecfr.gov/current/title-2/part-200/section-200.208}}.
\newblock
\newblock
\shownote{Specific Conditions}.


\bibitem[CFR(2024c)]%
        {CFR200211}
 \bibinfo{year}{2024}\natexlab{c}.
\newblock \bibinfo{title}{{2~CFR~§~200.211}}.
\newblock \bibinfo{howpublished}{\url{https://www.ecfr.gov/current/title-2/part-200/section-200.211}}.
\newblock
\newblock
\shownote{Information Contained in a Federal Award}.


\bibitem[CFR(2024c)]%
        {CFR200216}
 \bibinfo{year}{2024}\natexlab{c}.
\newblock \bibinfo{title}{{2~CFR~§~200.216}}.
\newblock \bibinfo{howpublished}{\url{https://www.ecfr.gov/current/title-2/part-200/section-200.216}}.
\newblock
\newblock
\shownote{Prohibition on Certain Telecommunications and Video Surveillance Equipment or Services}.


\bibitem[CFR(2024d)]%
        {CFR200300}
 \bibinfo{year}{2024}\natexlab{d}.
\newblock \bibinfo{title}{{2~CFR~§~200.300}}.
\newblock \bibinfo{howpublished}{\url{https://www.ecfr.gov/current/title-2/part-200/section-200.300}}.
\newblock
\newblock
\shownote{Statutory and National Policy Requirements}.


\bibitem[cfr(2024)]%
        {cfrrulemonitoring}
 \bibinfo{year}{2024}\natexlab{}.
\newblock \bibinfo{title}{2 CFR § 200.329}.
\newblock
\newblock
\urldef\tempurl%
\url{https://www.ecfr.gov/current/title-2/part-200/section-200.329}
\showURL{%
\tempurl}
\newblock
\shownote{Monitoring and reporting program performance}.


\bibitem[Bid(2024)]%
        {BidenHarrisAdministrationLaunches2024}
 \bibinfo{year}{2024}\natexlab{}.
\newblock \bibinfo{title}{The Biden-Harris Administration Launches the Federal Program Inventory to Make Federal Spending More Transparent and Accessible}.
\newblock
\newblock
\urldef\tempurl%
\url{https://www.whitehouse.gov/omb/briefing-room/2024/02/15/the-biden-harris-administration-launches-the-federal-program-inventory-to-make-federal-spending-more-transparent-and-accessible/}
\showURL{%
\tempurl}


\bibitem[Dep(2024)]%
        {DepartmentTransportationDiscretionary2024}
 \bibinfo{year}{2024}\natexlab{}.
\newblock \bibinfo{title}{Department of Transportation Discretionary Grants: Stakeholder Perspectives}.
\newblock
\newblock
\urldef\tempurl%
\url{https://transportation.house.gov/calendar/eventsingle.aspx?EventID=407248}
\showURL{%
\tempurl}


\bibitem[bip(2024)]%
        {bipartisansenateaiworkinggroupDrivingUSInnovation2024}
 \bibinfo{year}{2024}\natexlab{}.
\newblock \bibinfo{booktitle}{\emph{Driving U.S. Innovation In Artificial Intelligence}}.
\newblock \bibinfo{type}{{T}echnical {R}eport}. \bibinfo{institution}{Bipartisan Senate AI Working Group}.
\newblock
\urldef\tempurl%
\url{https://www.schumer.senate.gov/imo/media/doc/Roadmap_Electronic1.32pm.pdf}
\showURL{%
\tempurl}


\bibitem[FRA(2024)]%
        {FRAMEWORKNUCLEICACID2024}
 \bibinfo{year}{2024}\natexlab{}.
\newblock \bibinfo{booktitle}{\emph{Framework for Nucleic Acid Synthesis Screening}}.
\newblock \bibinfo{type}{{T}echnical {R}eport}. \bibinfo{institution}{National Science and Technology Council}.
\newblock
\urldef\tempurl%
\url{https://www.whitehouse.gov/wp-content/uploads/2024/04/Nucleic-Acid_Synthesis_Screening_Framework.pdf}
\showURL{%
\tempurl}


\bibitem[Ill(2024)]%
        {IllinoisProcurementCode2024}
 \bibinfo{year}{2024}\natexlab{}.
\newblock \bibinfo{title}{The Illinois Procurement Code}.
\newblock
\newblock
\urldef\tempurl%
\url{https://custom.statenet.com/public/resources.cgi?mode=show_text&id=ID:bill:IL2023000H5099&verid=IL2023000H5099_20240208_0_I&}
\showURL{%
\tempurl}


\bibitem[Pro(2024)]%
        {ProjectGrantsFY2024}
 \bibinfo{year}{2024}\natexlab{}.
\newblock \bibinfo{title}{Project Grants FY2024}.
\newblock
\newblock
\urldef\tempurl%
\url{https://www.usaspending.gov/search/?hash=ccb3c3c4a880d4717589b83093c4519c}
\showURL{%
\tempurl}


\bibitem[Uni(2024)]%
        {UniformGrantsGuidance2024}
 \bibinfo{year}{2024}\natexlab{}.
\newblock \bibinfo{title}{Uniform Grants Guidance 2024 Revision}.
\newblock
\newblock
\urldef\tempurl%
\url{https://www.cfo.gov/assets/files/Uniform%20Guidance%20_Reference%20Guides%20FINAL%204-2024.pdf}
\showURL{%
\tempurl}


\bibitem[App(2025)]%
        {AppendixPart200}
 \bibinfo{year}{2025}\natexlab{}.
\newblock \bibinfo{title}{Appendix I to Part 200, Title 2 -- Full Text of Notice of Funding Opportunity}.
\newblock
\newblock
\urldef\tempurl%
\url{https://www.ecfr.gov/current/title-2/appendix-Appendix%20I%20to%20Part%20200}
\showURL{%
\tempurl}


\bibitem[NIJ(2025)]%
        {NIJFY25Research2025}
 \bibinfo{year}{2025}\natexlab{}.
\newblock \bibinfo{title}{NIJ FY25 Research and Evaluation of Artificial Intelligence for Criminal Justice Purposes {\textbar} National Institute of Justice}.
\newblock
\newblock
\urldef\tempurl%
\url{https://nij.ojp.gov/funding/opportunities/o-nij-2025-172317}
\showURL{%
\tempurl}


\bibitem[Alder(2024)]%
        {USAgenciesPublish}
\bibfield{author}{\bibinfo{person}{Madison Alder}.} \bibinfo{year}{2024}\natexlab{}.
\newblock \showarticletitle{U.S. Agencies Publish Plans to Comply with White House AI Memo}.
\newblock  (\bibinfo{year}{2024}).
\newblock
\urldef\tempurl%
\url{https://fedscoop.com/u-s-agencies-publish-plans-to-comply-with-white-house-ai-memo/}
\showURL{%
\tempurl}


\bibitem[{Ali Crawford} and {Ido Wulkan}(2021)]%
        {alicrawfordFederalPrizeCompetitions2021}
\bibfield{author}{\bibinfo{person}{{Ali Crawford}} {and} \bibinfo{person}{{Ido Wulkan}}.} \bibinfo{year}{2021}\natexlab{}.
\newblock \bibinfo{booktitle}{\emph{Federal Prize Competitions: Using Competitions to Promote Innovation in Artificial Intelligence}}.
\newblock \bibinfo{type}{{T}echnical {R}eport}. \bibinfo{institution}{{Center for Security and Emerging Technology}}.
\newblock
\urldef\tempurl%
\url{https://cset.georgetown.edu/publication/federal-prize-competitions/}
\showURL{%
\tempurl}


\bibitem[Allen(2018)]%
        {allenFederalGrantPractice2018}
\bibfield{author}{\bibinfo{person}{Kenneth~J Allen}.} \bibinfo{year}{2018}\natexlab{}.
\newblock \bibinfo{booktitle}{\emph{Federal Grant Practice}}.
\newblock \bibinfo{publisher}{Thompson Reuters}.
\newblock


\bibitem[Andrews(2025)]%
        {andrews2025omb}
\bibfield{author}{\bibinfo{person}{Caitlin Andrews}.} \bibinfo{year}{2025}\natexlab{}.
\newblock \showarticletitle{OMB memos outline government's new AI use, procurement standards}.
\newblock \bibinfo{journal}{\emph{International Association of Privacy Professionals (IAPP)}} (\bibinfo{date}{9 April} \bibinfo{year}{2025}).
\newblock
\urldef\tempurl%
\url{https://iapp.org/news/a/omb-memos-outline-government-s-new-ai-use-procurement-standards}
\showURL{%
\tempurl}
\newblock
\shownote{Accessed: 2025-05-07}.


\bibitem[Arkhurst and Green~Williams(2024)]%
        {arkhurstRealizingJustice40Addressing2024}
\bibfield{author}{\bibinfo{person}{Bettina~K. Arkhurst} {and} \bibinfo{person}{Wyatt Green~Williams}.} \bibinfo{year}{2024}\natexlab{}.
\newblock \showarticletitle{Realizing Justice40: Addressing Structural Funding Barriers for Equitable Community Engagement in Energy RD\&D}.
\newblock \bibinfo{journal}{\emph{Journal of Science Policy \& Governance}} \bibinfo{volume}{23}, \bibinfo{number}{02} (\bibinfo{date}{March} \bibinfo{year}{2024}).
\newblock
\showISSN{2372-2193}
\urldef\tempurl%
\url{https://doi.org/10.38126/JSPG230201}
\showDOI{\tempurl}


\bibitem[Azoulay and Li(2020)]%
        {azoulayScientificGrantFunding2020}
\bibfield{author}{\bibinfo{person}{Pierre Azoulay} {and} \bibinfo{person}{Danielle Li}.} \bibinfo{year}{2020}\natexlab{}.
\newblock \bibinfo{booktitle}{\emph{Scientific Grant Funding}}.
\newblock \bibinfo{type}{{T}echnical {R}eport}. \bibinfo{institution}{National Bureau of Economic Research}, \bibinfo{address}{Cambridge, MA}.
\newblock
\urldef\tempurl%
\url{https://doi.org/10.3386/w26889}
\showDOI{\tempurl}


\bibitem[{Barry Friedman} et~al\mbox{.}(2023)]%
        {barryfriedmanFederalGovernmentsRole}
\bibfield{author}{\bibinfo{person}{{Barry Friedman}}, \bibinfo{person}{{Rachel Harmon}}, {and} \bibinfo{person}{{Farhang Heydari}}.} \bibinfo{year}{2023}\natexlab{}.
\newblock \showarticletitle{The Federal Government's Role in Local Policing}.
\newblock \bibinfo{journal}{\emph{Virginia Law Review}} \bibinfo{volume}{109}, \bibinfo{number}{8} (\bibinfo{year}{2023}).
\newblock
\urldef\tempurl%
\url{https://virginialawreview.org/articles/the-federal-governments-role-in-local-policing/}
\showURL{%
\tempurl}


\bibitem[Berry et~al\mbox{.}(2010)]%
        {berryPresidentDistributionFederal2010}
\bibfield{author}{\bibinfo{person}{Christopher~R. Berry}, \bibinfo{person}{Barry~C. Burden}, {and} \bibinfo{person}{William~G. Howell}.} \bibinfo{year}{2010}\natexlab{}.
\newblock \showarticletitle{The President and the Distribution of Federal Spending}.
\newblock \bibinfo{journal}{\emph{The American Political Science Review}} \bibinfo{volume}{104}, \bibinfo{number}{4} (\bibinfo{year}{2010}), \bibinfo{pages}{783--799}.
\newblock
\showISSN{0003-0554}
\showeprint[jstor]{40982897}
\urldef\tempurl%
\url{https://www.jstor.org/stable/40982897}
\showURL{%
\tempurl}


\bibitem[Biden(2023)]%
        {houseExecutiveOrderSafe2023}
\bibfield{author}{\bibinfo{person}{Joe Biden}.} \bibinfo{year}{2023}\natexlab{}.
\newblock \bibinfo{title}{Executive Order on the Safe, Secure, and Trustworthy Development and Use of Artificial Intelligence}.
\newblock
\newblock
\urldef\tempurl%
\url{https://web.archive.org/web/20250120132537/https://www.whitehouse.gov/briefing-room/presidential-actions/2023/10/30/executive-order-on-the-safe-secure-and-trustworthy-development-and-use-of-artificial-intelligence/}
\showURL{%
\tempurl}


\bibitem[Brown(1988)]%
        {brownFederalAgencysAuthority1988}
\bibfield{author}{\bibinfo{person}{Bruce~Allen Brown}.} \bibinfo{year}{1988}\natexlab{}.
\newblock \showarticletitle{A Federal Agency's Authority to Use a Cooperative Agreement for a Particular Project}.
\newblock \bibinfo{journal}{\emph{Public Contract Law Journal}} \bibinfo{volume}{18}, \bibinfo{number}{1} (\bibinfo{year}{1988}).
\newblock
\showISSN{0033-3441}
\urldef\tempurl%
\url{https://www.jstor.org/stable/25755554}
\showURL{%
\tempurl}


\bibitem[Buckland(2024)]%
        {bucklandNewWebScience2024}
\bibfield{author}{\bibinfo{person}{Francesca Buckland}.} \bibinfo{year}{2024}\natexlab{}.
\newblock \bibinfo{title}{New Web of Science Grants Index Helps Researchers Develop More Targeted Grant Proposals}.
\newblock
\newblock
\urldef\tempurl%
\url{https://clarivate.com/blog/new-web-of-science-grants-index-helps-researchers-develop-more-targeted-grant-proposals/}
\showURL{%
\tempurl}


\bibitem[{Cody Venzke}(2023)]%
        {ACLUEncouragesOMBa}
\bibfield{author}{\bibinfo{person}{{Cody Venzke}}.} \bibinfo{year}{2023}\natexlab{}.
\newblock \bibinfo{title}{ACLU Encourages OMB to Provide Robust Protections for Civil Rights and Civil Liberties in Government Uses of AI}.
\newblock
\newblock
\urldef\tempurl%
\url{https://www.aclu.org/documents/aclu-encourages-omb-to-provide-robust-protections-for-civil-rights-and-civil-liberties-in-government-uses-of-ai}
\showURL{%
\tempurl}


\bibitem[Coglianese(2023)]%
        {coglianeseProcurementArtificialIntelligence2023a}
\bibfield{author}{\bibinfo{person}{Cary Coglianese}.} \bibinfo{year}{2023}\natexlab{}.
\newblock \bibinfo{title}{Procurement and Artificial Intelligence}.
\newblock
\newblock
\urldef\tempurl%
\url{https://doi.org/10.2139/ssrn.4591724}
\showDOI{\tempurl}


\bibitem[Crump(2016)]%
        {crumpSurveillancePolicyMaking2016a}
\bibfield{author}{\bibinfo{person}{Catherine Crump}.} \bibinfo{year}{2016}\natexlab{}.
\newblock \showarticletitle{Surveillance Policy Making by Procurement}.
\newblock \bibinfo{journal}{\emph{Washington Law Review}} \bibinfo{volume}{91}, \bibinfo{number}{4} (\bibinfo{year}{2016}).
\newblock
\urldef\tempurl%
\url{https://doi.org/10.2139/ssrn.2737006}
\showDOI{\tempurl}


\bibitem[{Dan Bateyko} et~al\mbox{.}(2023)]%
        {danbateykoCDTCommentsOMB2023}
\bibfield{author}{\bibinfo{person}{{Dan Bateyko}}, \bibinfo{person}{{Hannah Quay-de la Vallee}}, \bibinfo{person}{{Ridhi Shetty}}, {and} \bibinfo{person}{{Alexandra Reeve Givens}}.} \bibinfo{year}{2023}\natexlab{}.
\newblock \bibinfo{booktitle}{\emph{CDT Comments on OMB Draft Guidance for Agency Use of AI}}.
\newblock \bibinfo{type}{{T}echnical {R}eport}. \bibinfo{institution}{Center for Democracy \& Technology}.
\newblock
\urldef\tempurl%
\url{https://cdt.org/insights/cdt-comments-on-omb-draft-guidance-for-agency-use-of-ai/}
\showURL{%
\tempurl}


\bibitem[for International~Development(2021)]%
        {GenderInequityArtificial2021}
\bibfield{author}{\bibinfo{person}{Agency for International~Development}.} \bibinfo{year}{2021}\natexlab{}.
\newblock \bibinfo{title}{Gender Inequity in Artificial Intelligence Activity}.
\newblock
\newblock
\urldef\tempurl%
\url{https://grants.gov/search-results-detail/332344}
\showURL{%
\tempurl}


\bibitem[{George W. Bush}(2002)]%
        {georgew.bushPresidentsManagementAgenda2002}
\bibfield{author}{\bibinfo{person}{{George W. Bush}}.} \bibinfo{year}{2002}\natexlab{}.
\newblock \bibinfo{booktitle}{\emph{The President's Management Agenda}}.
\newblock \bibinfo{type}{{T}echnical {R}eport}. \bibinfo{institution}{{Office of Management and Budget}}.
\newblock


\bibitem[{Grants}({[n.\,d.]})]%
        {granteligibilitygrantsgov}
\bibfield{author}{\bibinfo{person}{{Grants}}.} \bibinfo{year}{[n.\,d.]}\natexlab{}.
\newblock \bibinfo{title}{Grant Eligibility}.
\newblock
\newblock
\urldef\tempurl%
\url{https://www.grants.gov/learn-grants/grant-eligibility.html}
\showURL{%
\tempurl}


\bibitem[{Grants.gov}({[n.\,d.]})]%
        {GrantMakingAgenciesGrantsgov}
\bibfield{author}{\bibinfo{person}{{Grants.gov}}.} \bibinfo{year}{[n.\,d.]}\natexlab{}.
\newblock \bibinfo{title}{Grant-Making Agencies}.
\newblock
\newblock
\urldef\tempurl%
\url{https://www.grants.gov/learn-grants/grant-making-agencies/}
\showURL{%
\tempurl}


\bibitem[Grants.gov({[n.\,d.]})]%
        {OtherGrantMakingAgencies}
\bibfield{author}{\bibinfo{person}{Grants.gov}.} \bibinfo{year}{[n.\,d.]}\natexlab{}.
\newblock \bibinfo{title}{Other Grant-Making Agencies}.
\newblock
\newblock
\urldef\tempurl%
\url{https://www.grants.gov/learn-grants/grant-making-agencies/other-grant-making-agencies}
\showURL{%
\tempurl}


\bibitem[Grants.gov(2024)]%
        {XMLExtractGrantsgov}
\bibfield{author}{\bibinfo{person}{Grants.gov}.} \bibinfo{year}{2024}\natexlab{}.
\newblock \bibinfo{title}{XML Extract}.
\newblock
\newblock
\urldef\tempurl%
\url{https://grants.gov/xml-extract}
\showURL{%
\tempurl}


\bibitem[Hamburger(2021)]%
        {hamburgerPurchasingSubmissionConditions2021}
\bibfield{author}{\bibinfo{person}{Philip Hamburger}.} \bibinfo{year}{2021}\natexlab{}.
\newblock \bibinfo{booktitle}{\emph{Purchasing Submission: Conditions, Power, and Freedom}}.
\newblock \bibinfo{publisher}{Harvard University Press}, \bibinfo{address}{Cambridge, Massachusetts}.
\newblock
\showISBNx{978-0-674-25823-5}
\showLCCN{KF450.D85 H36 2021}


\bibitem[Henderson et~al\mbox{.}(2022)]%
        {hendersonAdsSequentialDecisionMaking2022a}
\bibfield{author}{\bibinfo{person}{Peter Henderson}, \bibinfo{person}{Ben Chugg}, \bibinfo{person}{Brandon Anderson}, {and} \bibinfo{person}{Daniel~E. Ho}.} \bibinfo{year}{2022}\natexlab{}.
\newblock \showarticletitle{Beyond Ads: Sequential Decision-Making Algorithms in Law and Public Policy}. In \bibinfo{booktitle}{\emph{Proceedings of the 2022 Symposium on Computer Science and Law}} \emph{(\bibinfo{series}{CSLAW '22})}. \bibinfo{publisher}{Association for Computing Machinery}, \bibinfo{address}{New York, NY, USA}, \bibinfo{pages}{87--100}.
\newblock
\showISBNx{978-1-4503-9234-1}
\urldef\tempurl%
\url{https://doi.org/10.1145/3511265.3550439}
\showDOI{\tempurl}


\bibitem[Keegan(2012)]%
        {keeganFederalGrantsinAidAdministration}
\bibfield{author}{\bibinfo{person}{Natalie Keegan}.} \bibinfo{year}{2012}\natexlab{}.
\newblock \bibinfo{booktitle}{\emph{Federal Grants-in-Aid Administration: A Primer}}.
\newblock \bibinfo{type}{{T}echnical {R}eport}. \bibinfo{institution}{Congressional Research Service}.
\newblock


\bibitem[Krafft et~al\mbox{.}(2020)]%
        {krafftDefiningAIPolicy2020}
\bibfield{author}{\bibinfo{person}{P.~M. Krafft}, \bibinfo{person}{Meg Young}, \bibinfo{person}{Michael Katell}, \bibinfo{person}{Karen Huang}, {and} \bibinfo{person}{Ghislain Bugingo}.} \bibinfo{year}{2020}\natexlab{}.
\newblock \showarticletitle{Defining AI in Policy versus Practice}. In \bibinfo{booktitle}{\emph{Proceedings of the AAAI/ACM Conference on AI, Ethics, and Society}}. \bibinfo{publisher}{ACM}, \bibinfo{address}{New York NY USA}, \bibinfo{pages}{72--78}.
\newblock
\showISBNx{978-1-4503-7110-0}
\urldef\tempurl%
\url{https://doi.org/10.1145/3375627.3375835}
\showDOI{\tempurl}


\bibitem[Lawrence et~al\mbox{.}(2023)]%
        {lawrenceBureaucraticChallengeAI2023b}
\bibfield{author}{\bibinfo{person}{Christie Lawrence}, \bibinfo{person}{Isaac Cui}, {and} \bibinfo{person}{Daniel Ho}.} \bibinfo{year}{2023}\natexlab{}.
\newblock \showarticletitle{The Bureaucratic Challenge to AI Governance: An Empirical Assessment of Implementation at U.S. Federal Agencies}. In \bibinfo{booktitle}{\emph{Proceedings of the 2023 AAAI/ACM Conference on AI, Ethics, and Society}} \emph{(\bibinfo{series}{AIES '23})}. \bibinfo{publisher}{Association for Computing Machinery}, \bibinfo{address}{New York, NY, USA}, \bibinfo{pages}{606--652}.
\newblock
\showISBNx{979-8-4007-0231-0}
\urldef\tempurl%
\url{https://doi.org/10.1145/3600211.3604701}
\showDOI{\tempurl}


\bibitem[{Lee-Easton} et~al\mbox{.}(2022)]%
        {lee-eastonUtilizationEvidencebasedIntervention2022a}
\bibfield{author}{\bibinfo{person}{Miranda~J. {Lee-Easton}}, \bibinfo{person}{Stephen Magura}, {and} \bibinfo{person}{Michael~J. Maranda}.} \bibinfo{year}{2022}\natexlab{}.
\newblock \showarticletitle{Utilization of Evidence-Based Intervention Criteria in U.S. Federal Grant Funding Announcements for Behavioral Healthcare}.
\newblock \bibinfo{journal}{\emph{INQUIRY: The Journal of Health Care Organization, Provision, and Financing}}  \bibinfo{volume}{59} (\bibinfo{date}{Jan.} \bibinfo{year}{2022}), \bibinfo{pages}{00469580221126295}.
\newblock
\showISSN{0046-9580}
\urldef\tempurl%
\url{https://doi.org/10.1177/00469580221126295}
\showDOI{\tempurl}


\bibitem[Levy et~al\mbox{.}(2021)]%
        {levyAlgorithmsDecisionMakingPublic2021a}
\bibfield{author}{\bibinfo{person}{Karen Levy}, \bibinfo{person}{Kyla~E. Chasalow}, {and} \bibinfo{person}{Sarah Riley}.} \bibinfo{year}{2021}\natexlab{}.
\newblock \showarticletitle{Algorithms and Decision-Making in the Public Sector}.
\newblock \bibinfo{journal}{\emph{Annual Review of Law and Social Science}} \bibinfo{volume}{17}, \bibinfo{number}{Volume 17, 2021} (\bibinfo{date}{Oct.} \bibinfo{year}{2021}), \bibinfo{pages}{309--334}.
\newblock
\showISSN{1550-3585, 1550-3631}
\urldef\tempurl%
\url{https://doi.org/10.1146/annurev-lawsocsci-041221-023808}
\showDOI{\tempurl}


\bibitem[MacMillan(2023)]%
        {EyesPoorCameras2023a}
\bibfield{author}{\bibinfo{person}{Douglas MacMillan}.} \bibinfo{year}{2023}\natexlab{}.
\newblock \showarticletitle{Eyes on the Poor: Cameras, Facial Recognition Watch over Public Housing}.
\newblock \bibinfo{journal}{\emph{Washington Post}} (\bibinfo{year}{2023}).
\newblock
\urldef\tempurl%
\url{https://www.washingtonpost.com/business/2023/05/16/surveillance-cameras-public-housing/}
\showURL{%
\tempurl}


\bibitem[McGarity(1993)]%
        {mcgarityPeerReviewAwarding}
\bibfield{author}{\bibinfo{person}{Thomas~O McGarity}.} \bibinfo{year}{1993}\natexlab{}.
\newblock \showarticletitle{Peer Review in Awarding Discretionary Grants in the Arts and Sciences}.
\newblock \bibinfo{journal}{\emph{Administrative Conference of the United States}} (\bibinfo{year}{1993}).
\newblock


\bibitem[Millar et~al\mbox{.}(2022)]%
        {millarTrendsUsePromotional2022}
\bibfield{author}{\bibinfo{person}{Neil Millar}, \bibinfo{person}{Bojan Batalo}, {and} \bibinfo{person}{Brian Budgell}.} \bibinfo{year}{2022}\natexlab{}.
\newblock \showarticletitle{Trends in the Use of Promotional Language (Hype) in National Institutes of Health Funding Opportunity Announcements, 1992-2020}.
\newblock \bibinfo{journal}{\emph{JAMA Network Open}} \bibinfo{volume}{5}, \bibinfo{number}{11} (\bibinfo{date}{Nov.} \bibinfo{year}{2022}), \bibinfo{pages}{e2243221}.
\newblock
\showISSN{2574-3805}
\urldef\tempurl%
\url{https://doi.org/10.1001/jamanetworkopen.2022.43221}
\showDOI{\tempurl}


\bibitem[Mulligan and Bamberger(2019)]%
        {mulliganProcurementPolicyAdministrative2019}
\bibfield{author}{\bibinfo{person}{Deirdre~K. Mulligan} {and} \bibinfo{person}{Kenneth~A. Bamberger}.} \bibinfo{year}{2019}\natexlab{}.
\newblock \showarticletitle{Procurement As Policy: Administrative Process for Machine Learning}.
\newblock \bibinfo{journal}{\emph{Berkeley Technology Law Journal}}  \bibinfo{volume}{34} (\bibinfo{date}{Oct.} \bibinfo{year}{2019}).
\newblock
\urldef\tempurl%
\url{https://doi.org/10.2139/ssrn.3464203}
\showDOI{\tempurl}


\bibitem[Nagle(1994)]%
        {nagleReviewEssentialsGrant1994}
\bibfield{author}{\bibinfo{person}{James~F. Nagle}.} \bibinfo{year}{1994}\natexlab{}.
\newblock \showarticletitle{Review of Essentials of Grant Law Practice}.
\newblock \bibinfo{journal}{\emph{Public Contract Law Journal}} \bibinfo{volume}{23}, \bibinfo{number}{2} (\bibinfo{year}{1994}), \bibinfo{pages}{301--303}.
\newblock
\showISSN{0033-3441}
\showeprint[jstor]{25754134}
\urldef\tempurl%
\url{https://www.jstor.org/stable/25754134}
\showURL{%
\tempurl}


\bibitem[Narayanan and Kapoor(2024)]%
        {narayananAISnakeOil2024a}
\bibfield{author}{\bibinfo{person}{Arvind Narayanan} {and} \bibinfo{person}{Sayash Kapoor}.} \bibinfo{year}{2024}\natexlab{}.
\newblock \bibinfo{booktitle}{\emph{AI Snake Oil: What Artificial Intelligence Can Do, What It Can't, and How to Tell the Difference}}.
\newblock \bibinfo{publisher}{Princeton University Press}, \bibinfo{address}{Princeton}.
\newblock
\showISBNx{978-0-691-24913-1}
\showLCCN{Q335 .N368 2024}


\bibitem[{Nestor Maslej, Loredana Fattorini, Raymond Perrault, Vanessa Parli, Anka Reuel, Erik Brynjolfsson, John Etchemendy, Katrina Ligett, Terah Lyons, James Manyika, Juan Carlos Niebles, Yoav Shoham, Russell Wald, and Jack Clark,}(2024)]%
        {nestormaslejloredanafattoriniraymondperraultvanessaparliankareuelerikbrynjolfssonjohnetchemendykatrinaligettterahlyonsjamesmanyikajuancarlosnieblesyoavshohamrussellwaldandjackclarkAIIndex20242024}
\bibfield{author}{\bibinfo{person}{{Nestor Maslej, Loredana Fattorini, Raymond Perrault, Vanessa Parli, Anka Reuel, Erik Brynjolfsson, John Etchemendy, Katrina Ligett, Terah Lyons, James Manyika, Juan Carlos Niebles, Yoav Shoham, Russell Wald, and Jack Clark,}}.} \bibinfo{year}{2024}\natexlab{}.
\newblock \bibinfo{booktitle}{\emph{The AI Index 2024 Annual Report}}.
\newblock \bibinfo{type}{{T}echnical {R}eport}. \bibinfo{institution}{Institute for Human-Centered AI}.
\newblock
\urldef\tempurl%
\url{https://aiindex.stanford.edu/wp-content/uploads/2024/04/HAI_2024_AI-Index-Report.pdf}
\showURL{%
\tempurl}


\bibitem[NIST(2018)]%
        {PSIAPPointCloud2018}
\bibfield{author}{\bibinfo{person}{NIST}.} \bibinfo{year}{2018}\natexlab{}.
\newblock \bibinfo{title}{PSIAP Point Cloud City}.
\newblock
\newblock
\urldef\tempurl%
\url{https://www.nist.gov/ctl/pscr/funding-opportunities/past-funding-opportunities/psiap-point-cloud-city}
\showURL{%
\tempurl}


\bibitem[of~Inspector General Department~of Homeland~Security(2021)]%
        {DHSHasMade}
\bibfield{author}{\bibinfo{person}{Office of~Inspector General Department~of Homeland~Security}.} \bibinfo{year}{2021}\natexlab{}.
\newblock \showarticletitle{DHS Has Made Progress in Meeting DATA Act Requirements, But Challenges Remain}.
\newblock  (\bibinfo{year}{2021}).
\newblock
\urldef\tempurl%
\url{https://www.oig.dhs.gov/reports/2020/dhs-has-made-progress-meeting-data-act-requirements-challenges-remain/oig-20-62-aug20}
\showURL{%
\tempurl}


\bibitem[of~the United~States(1982)]%
        {ResolvingDisputesFederal}
\bibfield{author}{\bibinfo{person}{Administrative~Conference of~the United~States}.} \bibinfo{year}{1982}\natexlab{}.
\newblock \bibinfo{title}{Resolving Disputes Under Federal Grant Programs}.
\newblock
\newblock
\urldef\tempurl%
\url{https://www.acus.gov/recommendation/resolving-disputes-under-federal-grant-programs}
\showURL{%
\tempurl}


\bibitem[of~the United~States(2023)]%
        {PeerReviewAwarda}
\bibfield{author}{\bibinfo{person}{Administrative~Conference of~the United~States}.} \bibinfo{year}{2023}\natexlab{}.
\newblock \bibinfo{title}{Peer Review in the Award of Discretionary Grants}.
\newblock
\newblock
\urldef\tempurl%
\url{https://www.acus.gov/document/peer-review-award-discretionary-grants}
\showURL{%
\tempurl}


\bibitem[on~Artificial~Intelligence(2021)]%
        {FinalReport}
\bibfield{author}{\bibinfo{person}{National Security~Commission on Artificial~Intelligence}.} \bibinfo{year}{2021}\natexlab{}.
\newblock \showarticletitle{Final Report}.
\newblock  (\bibinfo{year}{2021}).
\newblock


\bibitem[Pasachoff(2014)]%
        {pasachoffAgencyEnforcementSpendingb}
\bibfield{author}{\bibinfo{person}{Eloise Pasachoff}.} \bibinfo{year}{2014}\natexlab{}.
\newblock \showarticletitle{Agency Enforcement of Spending Clause Statutes: A Defense of the Funding Cut-Off}.
\newblock \bibinfo{journal}{\emph{Yale Law Journal}}  \bibinfo{volume}{124} (\bibinfo{date}{Nov.} \bibinfo{year}{2014}).
\newblock
\urldef\tempurl%
\url{https://www.yalelawjournal.org/article/agency-enforcement-of-spending-clause-statutes}
\showURL{%
\tempurl}


\bibitem[Pasachoff(2020)]%
        {pasachoffFederalGrantRules2020a}
\bibfield{author}{\bibinfo{person}{Eloise Pasachoff}.} \bibinfo{year}{2020}\natexlab{}.
\newblock \showarticletitle{Federal Grant Rules and Realities in the Intergovernmental Administrative State: Compliance, Performance, and Politics}.
\newblock \bibinfo{journal}{\emph{Yale Journal on Regulation}}  \bibinfo{volume}{37} (\bibinfo{year}{2020}).
\newblock


\bibitem[Pasachoff(2022)]%
        {pasachoffExecutiveBranchControl2022}
\bibfield{author}{\bibinfo{person}{Eloise Pasachoff}.} \bibinfo{year}{2022}\natexlab{}.
\newblock \showarticletitle{Executive Branch Control of Federal Grants: Policy, Pork, and Punishment}.
\newblock \bibinfo{journal}{\emph{Ohio State Law Journal}} \bibinfo{volume}{83}, \bibinfo{number}{6} (\bibinfo{date}{May} \bibinfo{year}{2022}).
\newblock
\urldef\tempurl%
\url{https://papers.ssrn.com/abstract=4107983}
\showURL{%
\tempurl}


\bibitem[Price(2019)]%
        {priceGrants2019}
\bibfield{author}{\bibinfo{person}{W.~Nicholson Price}.} \bibinfo{year}{2019}\natexlab{}.
\newblock \showarticletitle{Grants}.
\newblock \bibinfo{journal}{\emph{Berkeley Technology Law Journal}} \bibinfo{volume}{34}, \bibinfo{number}{1} (\bibinfo{year}{2019}), \bibinfo{pages}{1--66}.
\newblock
\showISSN{1086-3818}
\urldef\tempurl%
\url{https://www.jstor.org/stable/26755225}
\showURL{%
\tempurl}


\bibitem[{Quay-de la Vallee} et~al\mbox{.}(2024)]%
        {quay-delavalleeFederalGovernmentPower2024}
\bibfield{author}{\bibinfo{person}{Hannah {Quay-de la Vallee}}, \bibinfo{person}{Ridhi Shetty}, {and} \bibinfo{person}{Elizabeth Laird}.} \bibinfo{year}{2024}\natexlab{}.
\newblock \bibinfo{title}{The Federal Government's Power of the Purse: Enacting Procurement Policies and Practices to Support Responsible AI Use}.
\newblock
\newblock
\urldef\tempurl%
\url{https://cdt.org/insights/report-the-federal-governments-power-of-the-purse-enacting-procurement-policies-and-practices-to-support-responsible-ai-use/}
\showURL{%
\tempurl}


\bibitem[Raji et~al\mbox{.}(2022)]%
        {rajiFallacyAIFunctionality2022}
\bibfield{author}{\bibinfo{person}{Inioluwa~Deborah Raji}, \bibinfo{person}{I.~Elizabeth Kumar}, \bibinfo{person}{Aaron Horowitz}, {and} \bibinfo{person}{Andrew~D. Selbst}.} \bibinfo{year}{2022}\natexlab{}.
\newblock \showarticletitle{The Fallacy of AI Functionality}. In \bibinfo{booktitle}{\emph{2022 ACM Conference on Fairness, Accountability, and Transparency}}. \bibinfo{pages}{959--972}.
\newblock
\urldef\tempurl%
\url{https://doi.org/10.1145/3531146.3533158}
\showDOI{\tempurl}
\showeprint[arxiv]{2206.09511}~[cs]


\bibitem[Rylander(1998)]%
        {ScanwellChallengingPropriety}
\bibfield{author}{\bibinfo{person}{KM Rylander}.} \bibinfo{year}{1998}\natexlab{}.
\newblock \showarticletitle{"Scanwell" Plus: Challenging the Propriety of a Federal Agency's Decision to Use a Federal Grant and Cooperative Agreement}.
\newblock \bibinfo{journal}{\emph{Public Contract Law Journal}} \bibinfo{volume}{28}, \bibinfo{number}{1} (\bibinfo{year}{1998}), \bibinfo{pages}{69--87}.
\newblock


\bibitem[Schoeberl and Dohmen(2023)]%
        {schoeberlExaminingGovernmentGrant2023}
\bibfield{author}{\bibinfo{person}{Christian Schoeberl} {and} \bibinfo{person}{Hanna Dohmen}.} \bibinfo{year}{2023}\natexlab{}.
\newblock \showarticletitle{Spurring Science: Examining U.S. Government Grant Activity in AI}.
\newblock  (\bibinfo{date}{Nov.} \bibinfo{year}{2023}).
\newblock
\urldef\tempurl%
\url{https://cset.georgetown.edu/publication/spurring-science/}
\showURL{%
\tempurl}


\bibitem[Systems(2023)]%
        {CurrentStateGrants2023}
\bibfield{author}{\bibinfo{person}{REI Systems}.} \bibinfo{year}{2023}\natexlab{}.
\newblock \bibinfo{title}{Current State of Grants Management}.
\newblock
\newblock


\bibitem[{The White House}(2025)]%
        {whitehouse2025skinnybudget}
\bibfield{author}{\bibinfo{person}{{The White House}}.} \bibinfo{year}{2025}\natexlab{}.
\newblock \bibinfo{title}{The White House Office of Management and Budget Releases the President’s Fiscal Year 2026 Skinny Budget}.
\newblock \bibinfo{howpublished}{https://www.whitehouse.gov/briefings-statements/2025/05/the-white-house-office-of-management-and-budget-releases-the-presidents-fiscal-year-2026-skinny-budget/}.
\newblock
\newblock
\shownote{Press release}.


\bibitem[Trump(2025a)]%
        {trump2025ai}
\bibfield{author}{\bibinfo{person}{Donald~J. Trump}.} \bibinfo{year}{2025}\natexlab{a}.
\newblock \bibinfo{title}{Advancing Artificial Intelligence Education for American Youth}.
\newblock \bibinfo{howpublished}{\url{https://www.whitehouse.gov/presidential-actions/2025/04/advancing-artificial-intelligence-education-for-american-youth/}}.
\newblock
\newblock
\shownote{Executive Order 14277}.


\bibitem[Trump(2025b)]%
        {trump2025rescissions}
\bibfield{author}{\bibinfo{person}{Donald~J. Trump}.} \bibinfo{year}{2025}\natexlab{b}.
\newblock \bibinfo{title}{Initial Rescissions of Harmful Executive Orders and Actions}.
\newblock \bibinfo{howpublished}{https://www.whitehouse.gov/presidential-actions/2025/01/initial-rescissions-of-harmful-executive-orders-and-actions/}.
\newblock
\newblock
\shownote{Executive Order 14148}.


\bibitem[Trump(2025c)]%
        {trump2025barriersai}
\bibfield{author}{\bibinfo{person}{Donald~J. Trump}.} \bibinfo{year}{2025}\natexlab{c}.
\newblock \bibinfo{title}{Removing Barriers to American Leadership in Artificial Intelligence}.
\newblock \bibinfo{howpublished}{https://www.whitehouse.gov/presidential-actions/2025/01/removing-barriers-to-american-leadership-in-artificial-intelligence/}.
\newblock
\newblock
\shownote{Executive Order 14179}.


\bibitem[Vought(2020a)]%
        {voughtFiscalYearFY2020}
\bibfield{author}{\bibinfo{person}{Russell~T. Vought}.} \bibinfo{year}{2020}\natexlab{a}.
\newblock \bibinfo{title}{Fiscal Year (FY) 2022 Administration Research and Development Budget Priorities and Cross-Cutting Actions}.
\newblock
\newblock


\bibitem[Vought(2020b)]%
        {voughtM2106GuidanceRegulation2020}
\bibfield{author}{\bibinfo{person}{Russell~T. Vought}.} \bibinfo{year}{2020}\natexlab{b}.
\newblock \bibinfo{title}{M-21-06 Guidance for Regulation of Artificial Intelligence Applications}.
\newblock
\newblock
\urldef\tempurl%
\url{https://web.archive.org/web/20250106115446/https://www.whitehouse.gov/wp-content/uploads/2020/11/M-21-06.pdf}
\showURL{%
\tempurl}


\bibitem[Vought(2025)]%
        {vought2025}
\bibfield{author}{\bibinfo{person}{Russell~T. Vought}.} \bibinfo{year}{2025}\natexlab{}.
\newblock \bibinfo{booktitle}{\emph{Accelerating Federal Use of AI through Innovation, Governance, and Public Trust}}.
\newblock \bibinfo{type}{{T}echnical {R}eport} M-25-21. \bibinfo{institution}{Office of Management and Budget}.
\newblock
\urldef\tempurl%
\url{https://www.whitehouse.gov/wp-content/uploads/2025/02/M-25-21-Accelerating-Federal-Use-of-AI-through-Innovation-Governance-and-Public-Trust.pdf}
\showURL{%
\tempurl}
\newblock
\shownote{Director, Office of Management and Budget}.


\bibitem[Whittlestone and Clark(2021)]%
        {whittlestoneWhyHowGovernments2021}
\bibfield{author}{\bibinfo{person}{Jess Whittlestone} {and} \bibinfo{person}{Jack Clark}.} \bibinfo{year}{2021}\natexlab{}.
\newblock \bibinfo{title}{Why and How Governments Should Monitor AI Development}.
\newblock
\newblock
\showeprint[arxiv]{2108.12427}~[cs]
\urldef\tempurl%
\url{http://arxiv.org/abs/2108.12427}
\showURL{%
\tempurl}


\bibitem[Yeh(2017)]%
        {yehFederalGovernmentsAuthority2017}
\bibfield{author}{\bibinfo{person}{Brian~T. Yeh}.} \bibinfo{year}{2017}\natexlab{}.
\newblock \bibinfo{booktitle}{\emph{The Federal Government's Authority to Impose Conditions on Grant Funds}}.
\newblock \bibinfo{type}{{T}echnical {R}eport}. \bibinfo{institution}{Congressional Research Service}.
\newblock


\bibitem[Young(2024a)]%
        {youngAdvancingGovernanceInnovation2024}
\bibfield{author}{\bibinfo{person}{Shalanda~D Young}.} \bibinfo{year}{2024}\natexlab{a}.
\newblock \bibinfo{booktitle}{\emph{Advancing Governance, Innovation, and Risk Management for Agency Use of Artificial Intelligence}}.
\newblock \bibinfo{type}{{T}echnical {R}eport}. \bibinfo{institution}{{Office of Management and Budget}}.
\newblock


\bibitem[Young(2024b)]%
        {youngM2410MEMORANDUMHEADS}
\bibfield{author}{\bibinfo{person}{Shalanda~D Young}.} \bibinfo{year}{2024}\natexlab{b}.
\newblock \showarticletitle{M-24-10 Advancing Governance, Innovation, and Risk Management for Agency Use of Artificial Intelligence}.
\newblock  (\bibinfo{year}{2024}).
\newblock


\bibitem[Young(2024c)]%
        {youngAdvancingResponsibleAcquisition2024}
\bibfield{author}{\bibinfo{person}{Shalanda~D Young}.} \bibinfo{year}{2024}\natexlab{c}.
\newblock \bibinfo{title}{M-24-18: Advancing the Responsible Acquisition of Artificial Intelligence in Government}.
\newblock
\newblock


\end{thebibliography}

\appendix

\section{Dataset Selection}
\label{sec:comparison}
Returning to Grants.gov, we gathered short-length descriptions for the AI-related NOFOs in our dataset from the website’s daily metadata archive. These short-length descriptions vary from a few sentences to multiple pages. We searched the same AI keywords against each grant announcement's short-length description on Grants.gov. Our search found that \textbf{for our reviewed AI-related NOFOs, only 68/407 (~17\%) of Grants.gov summaries contained an AI-related keyword}.
We also review, as best we can, each grant program’s associated spending records. The Federal Funding Accountability and Transparency Act \cite{sen.coburnText2590109th2006} requires grant spending records be searchable on a single website, which is USASpending.gov. Yet tying NOFOs to their spending records is challenging. While both NOFOs and spending records contain matching identifiers called ``assistance listings,'' clerical errors result in multiple NOFOs sharing listings or a single NOFO filed under multiple listings. Using assistance listing codes from our AI grants for fiscal year 2024, we search all related records on USASpending (31,508 transactions) finding only 36 out of 55 responsive assistance listings, indicating \textbf{a third of spending records for our reviewed AI-related grant programs do not contain AI keywords}. These results should be interpreted cautiously; if a grant’s spending descriptions lack AI keywords, the grantee may simply not have developed or used an AI tool. 

Why might grant spending records not mention AI, even for opportunities that likely involve AI? One reason is that grantmakers provide detailed information announcements, while spending records tend to be shorter and less descriptive. Though publicly available, spending records are often too short or vague to answer whether a grant funds an AI use. In the illustrative case of HUD’s crime-fighting grants, nowhere in its spending records does HUD mention grantees purchased and used cameras or facial recognition technologies \cite{FederalAwardsSearch}. Such pinched descriptions are not unusual for spending records; a Government Accountability Office audit of the Department of Homeland Security (DHS)’s USASpending.gov records found that the agency used jargon and shorthand which made spending records unintelligible \cite{DHSHasMade}. 
\begin{table}[h]
\caption{List of Abbreviations and Corresponding Agency Names}
\label{table:agency_abbreviations}
\begin{tabular}{ll}
\toprule
\textbf{Abbreviation} & \textbf{Agency Name} \\
\midrule
AC & Americorps \\
ARPAH & Advanced Research Projects Agency for Health \\
CPSC & Consumer Product Safety Commission \\
CNCS & Corporation for National and Community Service \\
DC & Denali Commission \\
DHS & Department of Homeland Security \\
DOC & Department of Commerce \\
DOE & Department of Energy \\
DOI & Department of the Interior \\
DOL & Department of Labor \\
DOS & Department of State \\
DOT & Department of Transportation \\
ED & Department of Education \\
EAC & Election Assistance Commission \\
EPA & Environmental Protection Agency \\
FCC & Federal Communications Commission \\
FDA & Food and Drug Administration \\
FMCS & Federal Mediation and Conciliation Service \\
HHS & Department of Health and Human Services \\
HUD & Department of Housing and Urban Development \\
IAF & Inter-American Foundation \\
IMLS & Institute of Museum and Library Services \\
LOC & Library of Congress \\
MCC & Millennium Challenge Corporation \\
NASA & National Aeronautics and Space Administration \\
NCD & National Council on Disability \\
NCUA & National Credit Union Administration \\
NEA & National Endowment for the Arts \\
NEH & National Endowment for the Humanities \\
NRC & Nuclear Regulatory Commission \\
ONDCP & Office of National Drug Control Policy \\
SBA & Small Business Administration \\
SCRC & Southeast Crescent Regional Commission \\
SSA & Social Security Administration \\
USAID & United States Agency for International Development \\
USDA & United States Department of Agriculture \\
USDOT & United States Department of the Treasury \\
USDOJ & United States Department of Justice \\
VA & Department of Veterans Affairs \\
\bottomrule
\end{tabular}
\end{table}
\label{fig:agencynames}
In spite of incomplete spending records, studies analyzing AI-related grant spending still provide great insights, often by relying on data processed and aggregated by private analytics companies. Companies like Clarivate \cite{bucklandNewWebScience2024}, Govini \cite{GoviniArkPlatform}, and Dimensions.ai \cite{DimensionsAIMost} gather grant data from across federal agencies, categorizing and connecting data to help researchers uncover trends. Researchers like those at Stanford’s Center Human-Centered Artificial Intelligence and Georgetown’s Center for Security and Emerging Technology effectively use these companies’ products to document AI grant spending patterns \cite{nestormaslejloredanafattoriniraymondperraultvanessaparliankareuelerikbrynjolfssonjohnetchemendykatrinaligettterahlyonsjamesmanyikajuancarlosnieblesyoavshohamrussellwaldandjackclarkAIIndex20242024,schoeberlExaminingGovernmentGrant2023}. However, these tools have their limitations. Companies can process, link, and predict to identify AI spending, but cannot fill in what is missing from a flawed spending record. Although companies do not always disclose their data sources, most appear to rely on the same incomplete federal sources. Further, the lack of transparency around data sources and subscription fees create replication problems and financial barriers to entry. 

Another potential data source are grant agencies’ voluntary disclosures, but these also vary in quality and are not centralized. For those agencies that chose to disclose additional information, they are not required to publish, archive, or update such disclosures with any regularity. Researchers may request grant-related documents through a Freedom of Information Act request, but doing so is time-intensive and costly. In some cases, that information may not even be FOIA eligible \cite{allenFederalGrantPractice2018}. 

These results suggest that no one dataset provides enough detail to identify all federal assistance that supports AI use. Government efforts to make federal spending more transparent serve to underscore the challenge of improving grant record quality; the newly minted Federal Program Inventory for grants, for example, relies on the same, often faulty upstream data source, USASpending.gov \cite{BidenHarrisAdministrationLaunches2024}. To improve on this status quo, researchers may consider combining public datasets when investigating federal assistance.

\section{Data Gathering and Coverage}
\label{sec:coverage}

\begin{figure*}[t]
 \centering
 \includegraphics[width=\textwidth, height=\textheight]{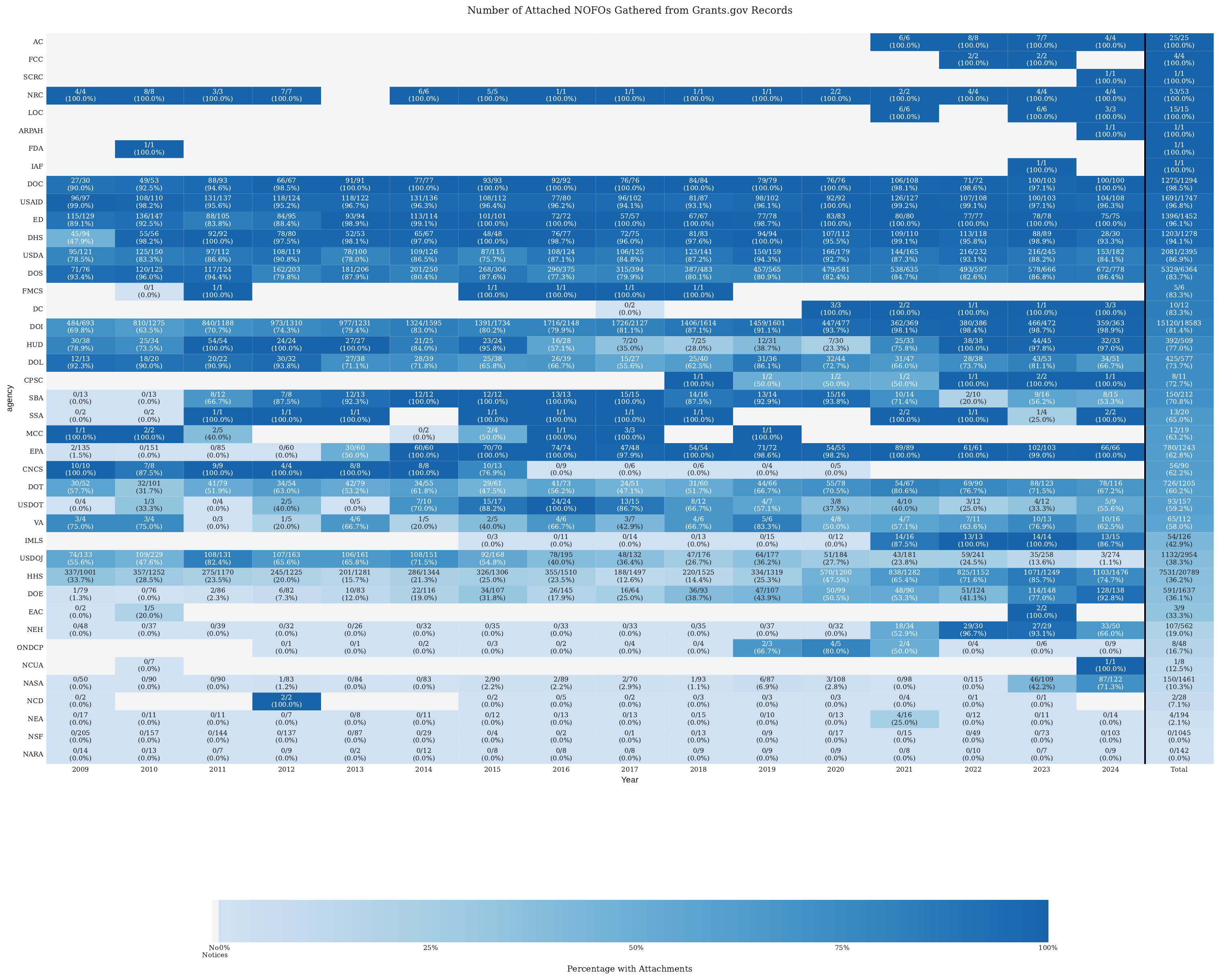}
 \caption{The above heatmap shows our Notice of Funding Opportunity collection from Grants.gov across federal agencies and years. Each cell displays both raw counts (attachments/total notices) and percentages. Cell colors indicate the percentage of notices with successfully collected attachments, ranging from light to dark blue. Light grey cells indicate years with no grant notices for that agency.}
 \label{fig:agency_attachments}
\end{figure*}

Grants.gov contains more than just grants; the database includes records of notices of intent, procurement contracts, partnership intermediary requests, requests for statements of interest, and requests for information. Grants.gov does not offer a public API for gathering NOFOs specifically; instead we use an exposed API that allows us to gather attachments, numbered sequentially from the oldest record to the newest. We subset to NOFOs by linking attachments to Grants.gov-provided metadata by their Opportunity Numbers, a unique identifier found in the zip filename. Although we take care to include all NOFOs attached to grant records in our study, some NOFOs cannot be readily associated with their Grants.gov record, like those that have been mislabeled. As a result, the total count is an underestimate of voluntary NOFO attachments.

\end{document}